\documentclass[unnumsec,webpdf,contemporary,large]{oup-authoring-template}%

\graphicspath{{Fig/}}

\usepackage{graphicx}
\usepackage{latexsym}
\usepackage{booktabs}
\usepackage{amsmath}
\usepackage{subcaption}
\usepackage{url}
\usepackage[hidelinks]{hyperref}
\usepackage{multirow}
\usepackage{multicol}
\usepackage{bm}
\usepackage{wrapfig}
\usepackage{amssymb}
\usepackage{pifont}
\usepackage{xcolor}
\newcommand{\cmark}{\ding{51}}%
\newcommand{\xmark}{\ding{55}}%

\theoremstyle{thmstyleone}%
\theoremstyle{thmstyletwo}%
\theoremstyle{thmstylethree}%

\begin{document}

\journaltitle{Journal Title Here}
\DOI{DOI HERE}
\copyrightyear{2022}
\pubyear{2019}
\access{Advance Access Publication Date: Day Month Year}
\appnotes{Paper}

\firstpage{1}


\title[SSM-DTA]{Breaking the Barriers of Data Scarcity in Drug-Target Affinity Prediction}

\author[1,4]{Qizhi Pei\ORCID{0000-0002-7242-422X}}
\author[2,$\ast$]{Lijun Wu\ORCID{0000-0002-3530-590X}}
\author[3]{Jinhua Zhu\ORCID{0000-0003-2157-9077}}
\author[2]{Yingce Xia\ORCID{0000-0001-9823-9033}}
\author[1]{Shufang Xie\ORCID{0000-0002-7126-0139}}
\author[2]{Tao Qin\ORCID{0000-0002-9095-0776}}
\author[2]{Haiguang Liu\ORCID{0000-0001-7324-6632}}
\author[2]{Tie-Yan Liu}
\author[1,5,$\ast$]{Rui Yan}

\authormark{Qizhi et al.}

\address[1]{\orgdiv{Gaoling School of Artificial Intelligence}, \orgname{Renmin University of China}, \orgaddress{\street{No.59, Zhong Guan Cun Avenue, Haidian District}, \postcode{100872}, \state{Beijing}, \country{China}}}
\address[2]{\orgname{Microsoft Research AI4Science}, \orgaddress{\street{No.5, Dan Ling Street, Haidian District}, \postcode{100080}, \state{Beijing}, \country{China}}}
\address[3]{\orgdiv{CAS Key Laboratory of GIPAS, EEIS Department}, \orgname{University of Science and Technology of China}, \orgaddress{\street{No.96, JinZhai Road, Baohe District}, \postcode{230026}, \state{Hefei, Anhui Province}, \country{China}}}
\address[4]{\orgname{Engineering Research Center of Next-Generation Intelligent Search and Recommendation, Ministry of Education}}
\address[5]{\orgname{Beijing Key Laboratory of Big Data Management and Analysis Methods}}

\corresp[$\ast$]{Corresponding authors:
Lijun Wu, Microsoft Research AI4Science,
Beijing, China, E-mail: \href{email:lijuwu@microsoft.com}{lijuwu@microsoft.com};
Rui Yan, Gaoling School of Artificial Intelligence, Renmin University of China, Beijing, China, E-mail: \href{email:ruiyan@ruc.edu.cn}{ruiyan@ruc.edu.cn}}

\received{Date}{0}{Year}
\revised{Date}{0}{Year}
\accepted{Date}{0}{Year}



\abstract{
Accurate prediction of Drug-Target Affinity (DTA) is of vital importance in early-stage drug discovery, facilitating the identification of drugs that can effectively interact with specific targets and regulate their activities. While wet experiments remain the most reliable method, they are time-consuming and resource-intensive, resulting in limited data availability that poses challenges for deep learning approaches. Existing methods have primarily focused on developing techniques based on the available DTA data, without adequately addressing the data scarcity issue. To overcome this challenge, we present the SSM-DTA framework, which incorporates three simple yet highly effective strategies: (1) A multi-task training approach that combines DTA prediction with masked language modeling (MLM) using paired drug-target data. (2) A semi-supervised training method that leverages large-scale unpaired molecules and proteins to enhance drug and target representations. This approach differs from previous methods that only employed molecules or proteins in pre-training. (3) The integration of a lightweight cross-attention module to improve the interaction between drugs and targets, further enhancing prediction accuracy. Through extensive experiments on benchmark datasets such as BindingDB, DAVIS, and KIBA, we demonstrate the superior performance of our framework. Additionally, we conduct case studies on specific drug-target binding activities, virtual screening experiments, drug feature visualizations, and real-world applications, all of which showcase the significant potential of our work. In conclusion, our proposed SSM-DTA framework addresses the data limitation challenge in DTA prediction and yields promising results, paving the way for more efficient and accurate drug discovery processes. Our code is available at \url{https://github.com/QizhiPei/SSM-DTA}.
}

\boxedtext{
\begin{itemize}
    \item We propose a semi-supervised multi-task training framework with simple yet effective strategies for DTI prediction to alleviate the data limitation issue. 
    \item We demonstrate the state-of-the-art performance of the SSM-DTA framework by conducting multiple experiments on DTI benchmark datasets. 
    \item We provide extensive studies to show the potential of SSM-DTA, e.g., it can well capture the structural knowledge with accurate binding information between drugs and targets. 
\end{itemize}
}

\maketitle

\section{Introduction}
As healthcare management continues to advance, the research of drugs and their various applications, such as Drug-Drug Interactions (DDI), drug repurposing, drug synergy prediction, and Drug-Target Interactions (DTI), has become increasingly important. Among these areas, the prediction of Drug-Target Affinity (DTA) is particularly crucial in drug research\cite{paul2010improve}, as it involves accurately forecasting the binding effect of a drug to a protein target. While computational methods like molecular docking~\cite{trott2010autodock} and molecular dynamics simulations~\cite{liu2018molecular} can be effective for predicting DTA, they can be costly due to their high computational demands.

Deep learning has been promising in recent years to predict drug-target affinity (DTA) due to its high accuracy and efficiency~\cite{tsubaki2019compound,karimi2019deepaffinity,huang2020deeppurpose,chen2020transformercpi}. However, the effectiveness of deep learning models relies heavily on the availability of large amounts of labeled training data. Unfortunately, there is a limited amount of labeled binding affinity data available for recognized drugs and targets, as collecting this data through experiments can be costly and time-consuming. For example, the commonly used BindingDB dataset contains approximately 2 million binding data points, but as noted in~\cite{karimi2019deepaffinity}, the data is of low quality and the affinity labels for the same compound-protein pairs are often noisy due to various factors such as variations in experimental conditions and different data sources. After following the same data filtration protocols as~\cite{karimi2019deepaffinity}, the remaining clean data only contains around $200k$ points, which is small compared to datasets in natural language processing and computer vision (which often contain millions to billions of data points) where deep learning has seen great success. This limited amount of labeled data poses a challenge for deep learning-based DTA prediction.

To address the issue of drug-target affinity (DTA) prediction, we present the Semi-Supervised Multi-task Training (SSM) framework. This framework includes three strategies to improve DTA prediction accuracy:
(1) We conduct multi-task training on labeled drug-target pairs. Besides the DTA prediction task, we incorporate a masked language modeling (MLM) task~\cite{devlin2018bert} on both drugs and targets, which can effectively improve DTA prediction accuracy.
(2) In addition, inspired by the success of self-supervised learning~\cite{liu2019roberta,devlin2018bert}, we leverage large-scale unlabeled molecule and protein data to help with drug and target representation learning. We find that common pre-training (on unlabeled data) and fine-tuning (on supervised data) paradigm can not work well in DTA prediction, since the separate pre-training on unlabeled molecule and protein ignores the importance of interaction between paired drug and target when fine-tuning. Therefore, we propose a different method to mix unlabeled data together with paired data and perform semi-supervised training for DTA prediction.
(3) We introduce an efficient cross-attention module to explicitly capture interaction information between drugs and targets, further improving DTA prediction accuracy.

We evaluate SSM-DTA framework by conducting experiments on several benchmarks, including the BindindDB, DAVIS, and KIBA. The results showed that SSM-DTA significantly improved the accuracy of DTA prediction, for example, a large reduction in root-mean-square-error (RMSE) on the BindingDB IC$_{50}$ dataset, a $5\%$ improvement over state-of-the-art methods. SSM-DTA also showed strong generalization ability on unknown drugs. In addition, case studies of drug-target binding activities showed that the model was able to accurately identify important atomic groups and amino acids within binding sites. The model's ability to group drugs by their targets was also visualized, revealing good performance in this regard. Real-world applications of SSM-DTA framework to detect targets for drugs also demonstrated its strong generalization capabilities. Overall, these results highlight the potential of the SSM framework in DTA research.
The main contributions are:
\begin{itemize}
    \item We propose a semi-supervised multi-task training framework with simple yet effective strategies for DTI prediction to alleviate the data limitation issue. 
    \item We demonstrate the state-of-the-art performance of the SSM-DTA framework by conducting multiple experiments on DTI benchmark datasets. 
    \item We provide extensive studies to show the potential of SSM-DTA, e.g., it can well capture the structural knowledge with accurate binding information between drugs and targets. 
\end{itemize}

\section{Related Work}
\subsection{Drug-Target Interaction/Affinity Prediction}
The DTI/DTA prediction methods can be classified into structure-based and structure-free approaches. 
For structure-based methods, molecular docking~\cite{luo2016molecular} and molecular dynamics (MD) simulations~\cite{liu2018molecular} are typical works, which require the 3D structure inputs of drug and protein and utilize predefined force fields to estimate the binding affinity at the atomic level. 
However, the strong dependency on high-quality 3D structures comes with a major limitation. 
Structure-free predictions are then developed. Similarity-based methods are popular, which calculate the similarity metrics as descriptors for drugs and targets to predict DTA~\cite{cichonska2017computational}, e.g., KronRLS~\cite{pahikkala2015toward} and SimBoost~\cite{he2017simboost}. Recently, deep-learning-based methods are promising, such as DeepAffinity~\cite{karimi2019deepaffinity}, MONN~\cite{li2020monn}, GeneralizedDTA~\cite{lin2022generalizeddta}, FusionDTA~\cite{yuan2022fusiondta}, and others~\cite{tsubaki2019compound,ozturk2018deepdta,li2022bacpi,nguyen2022mitigating}, which usually adopt different networks (e.g., CNN, RNN, Transformers) on the labeled data for affinity prediction. DeepAffinity~\cite{karimi2019deepaffinity} employs unsupervised pre-training on a seq2seq RNN model using compound SMILES or protein SPS representations.
GeneralizedDTA~\cite{lin2022generalizeddta} is designed for unknown drug discovery, which combines self-supervised pre-training and multi-task learning with a dual adaptation mechanism to enhance protein and drug feature representation and improve generalization capability.
FusionDTA~\cite{yuan2022fusiondta} uses a multi-head linear attention mechanism for feature aggregation and knowledge distillation to reduce parameter redundancy, resulting in improved performance and efficiency in the DTA task.
~\cite{nguyen2022mitigating} uses transfer learning from chemical–chemical interaction (CCI) and protein-protein interaction (PPI) tasks to enhance drug-target interaction predictions.

\subsection{Molecule and Protein Pre-training}
For molecule pre-training methods, there are two rough types based on molecule representations: the sequential model pre-train~\cite{fabian2020molecular} based on the simplified molecular-input line-entry system (SMILES)~\cite{weininger1988smiles} and the graph neural network (GNN) pre-train~\cite{hu2019strategies} based on the molecule graph structures. The SMILES-based pre-train typically takes the Transformer network, motivated by its excellent performances demonstrated on related fields, such as SMILES-BERT~\cite{wang2019smiles} and Chemberta~\cite{chithrananda2020chemberta}. Different GNN models are adopted for graph structure pre-train, and the pre-train tasks include masked training that are performed on different graph parts, such as N-gram~\cite{liu2019n}, AttrMasking~\cite{hu2019strategies}, ContextPredict~\cite{hu2019strategies}, MotifPredict~\cite{rong2020self}, and also the contrastive learning based methods~\cite{qiu2020gcc}.
Protein pre-train depends on the protein sequences, which are fed into Transformer encoder for pre-training~\cite{brandes2021proteinbert}. Representative works include TAPE~\cite{rao2019evaluating} and ESM~\cite{rives2021biological}. The above works all take protein or molecule only for pre-train, without considering the relationship between them. 

\begin{figure*}[h]
    \centering
    \includegraphics[scale=0.7]{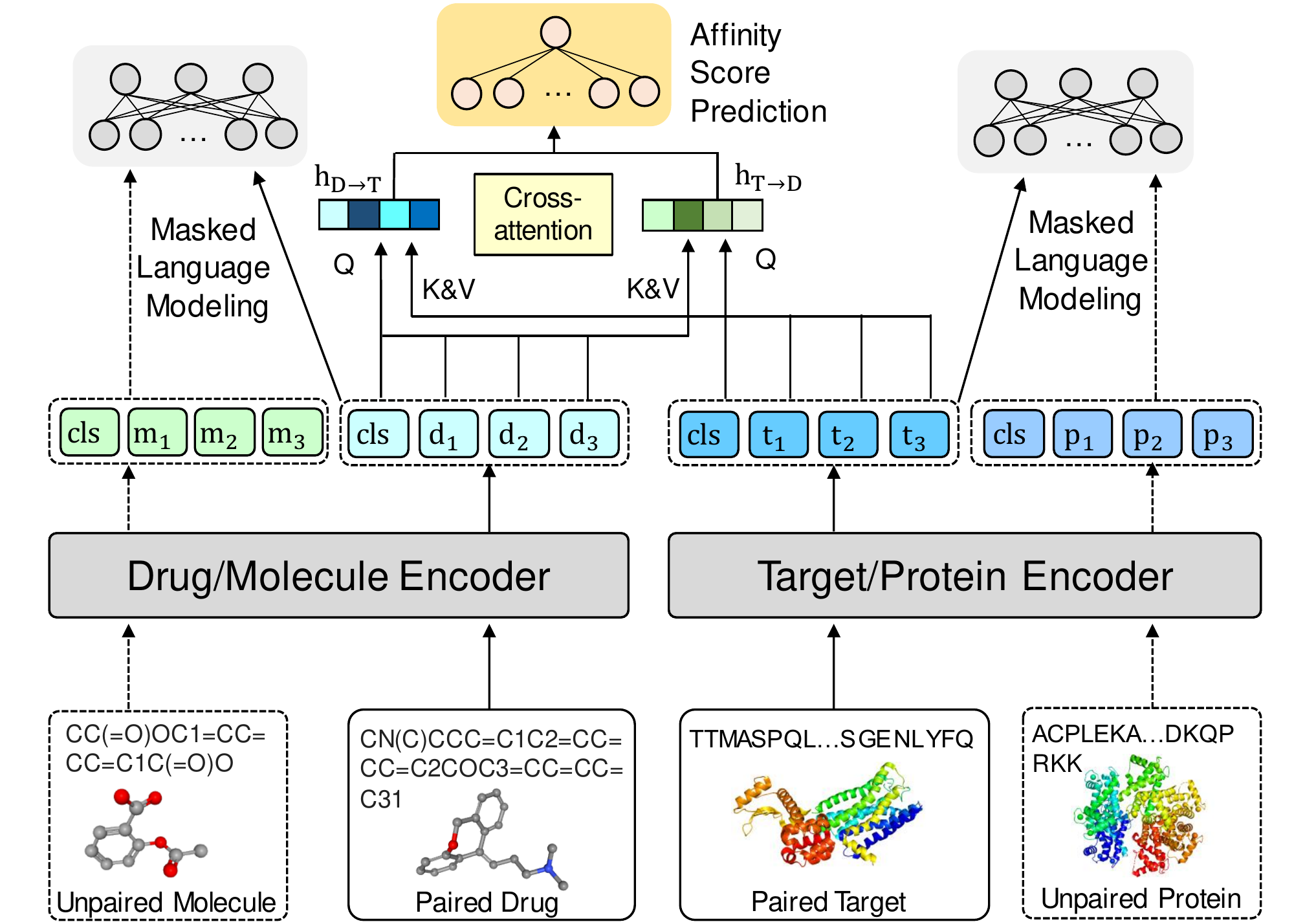}
    \caption{The overall framework of our SSM-DTA. The two encoders are Transformers and take the drug/molecule and target/protein sequences as input respectively. The MLM (Masked Language Modeling) training is conducted on both unlabeled (unpaired) data and paired data. The cross-attention module is used between drug-target paired data, where Q is a query, K\&V are key and value.}
    \label{fig:arch}
\end{figure*}

\section{Methods}
\noindent{\bf{Preliminary}}
We first give the necessary notations and definitions of DTA prediction.
Given a DTA dataset $\mathcal{DT} = \{(D, T, y)_i\}_{i=1}^N$, where $(D, T, y)$ is the triplet of DTA data sample, $N$ is the sample size, $D$ is one drug from a drug dataset, $T$ is one target from a target dataset, label $y$ (a floating number) is the affinity for the $(D, T)$ pair. 
DTA prediction is a regression task to predict the binding affinity score between the drug and target, learning a mapping function $\mathcal{F}: D \times T \to y$. 
The affinity score (label $y$) is a quantitative measure of the strength of binding between a drug and target. 
We take the SMILES string as the representation of drugs, which is a sequence resulting from traversing the molecule graph using depth-first search and some specific rules, e.g., $D=\{d_i\}_{i=1}^{\vert D \vert}$, where $\vert D \vert$ is the length of SMILES and $d_i$ is the token in the string. For a target $T$, we use the FASTA sequence of the protein, which consists of amino acid tokens, e.g., $T=\{t_i\}_{i=1}^{\vert T \vert}$, $\vert T \vert$ is the length and $t_i$ is the amino acid token. Similarly, the unlabeled molecule sequence is denoted as $M=\{m_i\}_{i=1}^{\vert M \vert}$, and the protein sequence is denoted as $P=\{p_i\}_{i=1}^{\vert P \vert}$.

The overall architecture of the SSM-DTA framework is illustrated in Figure~\ref{fig:arch}.
Our training data consists of three parts: paired drug-target data, unlabeled molecule data, and unlabeled protein sequences. The molecule and drug sequences are processed by a Transformer model~\cite{vaswani2017attention} $\mathcal{M_D}$, and the protein and target sequences are processed by another Transformer model $\mathcal{M_T}$\footnote{The detailed encoding process of Transformers $\mathcal{M_D}$ and $\mathcal{M_T}$ is introduced in the Appendix.}.
For the paired data $(D, T)$, the model performs a regression task for DTA prediction and also undergoes MLM training. After encoding the drug and target sequences and obtaining the attended representations $h_{D\to T}$ and $h_{T\to D}$ through the cross-attention module, they are concatenated together for the final affinity prediction with an MLP prediction head. The unlabeled molecule and protein data are only used for MLM training.

Before feeding the paired drug and target sequences into $\mathcal{M_D}$ and $\mathcal{M_T}$, we first add a special \texttt{[cls]} token at the beginning of the sequences. Specifically, the drug $D$ and target $T$ are formed as $D = \{\texttt{[cls]}_D, \{d_i\}_{i=1}^{\vert D \vert}\}$ and $T = \{\texttt{[cls]}_T,\{t_i\}_{i=1}^{\vert T \vert}\}$. These sequences are then encoded by $\mathcal{M_D}$ and $\mathcal{M_T}$ to obtain hidden states $H_D = \{h_{\texttt{[cls]}_D}, \{h_{d_i}\}_{i=1}^{\vert D \vert}\}$ and $H_T = \{h_{\texttt{[cls]}_T}, \{h_{t_i}\}_{i=1}^{\vert T \vert}\}$. The same operation is performed on unlabeled $M$ and $P$.

\noindent{\textbf{Interaction with Cross-attention Module}}
DTA prediction aims to predict the binding affinity resulting from the \emph{interaction} between drug and target. Therefore it is important to model the interaction effectively. 
Recent \cite{li2020monn,karimi2019deepaffinity} have focused on the design of the interaction module and introduced the attention mechanism~\cite{niu2021review}. They use joint attention that operates on each token of drug and target sequences, resulting in a pairwise interaction attention matrix over all tokens, which requires high computational cost.

We introduce a lightweight cross-attention module, which is much simpler than the pairwise interaction mechanism. To calculate our cross-attention, we use the $h_{\texttt{[cls]}_D}$ and $h_{\texttt{[cls]}_T}$ elements of $H_D$ and $H_T$ to represent the drug $D$ and target $T$ respectively.
Cross-attention is then performed between the \texttt{[cls]} tokens and the paired sequences, where the query is $h_{\texttt{[cls]}_D}$ or $h_{\texttt{[cls]}_T}$, and the key, value are $H_T$ or $H_D$:
\begin{align}
\small
    h_{D\to T} &= \texttt{softmax} (\frac{(h_{\texttt{[cls]}_D} W_{1}) (H_T W_2)^{\mathrm{T}}}{\sqrt{d}}) (H_T W_3), \nonumber \\
    h_{T\to D} &= \texttt{softmax} (\frac{(h_{\texttt{[cls]}_T} W_{4}) (H_D W_5)^{\mathrm{T}}}{\sqrt{d}}) (H_D W_6), \nonumber
\end{align}
where $d$ is the hidden dimension, and $W$s are parameter matrices. After the cross-attention, the attended representations are then concatenated and fed into an MLP head for DTA prediction.
\begin{equation}
\small
    y'=\texttt{MLP}(([h_{D\to T}, h_{T\to D}]), \nonumber
\end{equation}
where MLP means multilayer perceptron, $[]$ means concatenation, and $y'$ is the predicted binding affinity. The loss function for DTA task is the mean squared error (MSE):
\begin{equation}
    \mathcal{L}_{mse} = (y - y')^2, \nonumber
\end{equation}
where $y$ is the ground truth binding affinity.

\subsection{Multi-task Training with MLM}
Since we are targeting alleviating the data limitation issue of DTA prediction, multi-task training~\cite{liu2016recurrent,min2021pre} is a preferred method that is widely adopted. In this work, inspired by the general self-supervised methods that take the masked language modeling (MLM) objective for pre-training~\cite{devlin2018bert,liu2019roberta}, we take the same spirit in our DTA prediction. Besides the DTA prediction, we introduce MLM to form a multi-task training. Note that both DTA prediction and MLM are conducted upon the original paired drug-target data.

For a drug $D=\{d_i\}_{i=1}^{\vert D \vert}$ and target $T=\{t_i\}_{i=1}^{\vert T \vert}$, we follow~\cite{devlin2018bert} to randomly replace some of the tokens $d_i$, $t_i$ in the sequence by a special \texttt{[MASK]} token with some mask ratios~\cite{devlin2018bert}, and then try to recover the original sequence by the masked sequences $D'$ and $T'$,
\begin{equation}
    \label{eqn:mlm_pair}
    \mathcal{L}_{mlm}^D=-\sum_{k=1}^{M_D}\log P(d_k\|D'),  \mathcal{L}_{mlm}^T=-\sum_{k=1}^{M_T}\log P(t_k\|T'), \nonumber
\end{equation}
where $d_k$, $t_k$ are the masked tokens, $M_D$, $M_T$ are the masked token numbers in drug and target sequences,
$log$ is log-likelihood, and $P$ means the conditional probability. Both ${L}_{mlm}^D$ and ${L}_{mlm}^T$ are based on the negative log-likelihood of the masked tokens and the goal is to predict the original tokens from the masked sequences.
During training, the MLM objectives are jointly optimized with DTA prediction to enhance drug and target representations. In this way, though we do not incorporate more labeled DTA data, we maximize the potential of the original DTA data.

\subsection{Semi-supervised Training with Large-scale Unlabeled Molecule and Protein Data}
Pre-training on large amounts of unlabeled data in a self-supervised manner has shown promising results in natural language processing~\cite{devlin2018bert,liu2019roberta}, image processing~\cite{caron2019unsupervised}, and video processing~\cite{liu2021video}. In the chemistry and biology domains, pre-trained models have also demonstrated success~\cite{hu2019strategies}.
In this work, we also leverage large amounts of unlabeled molecules and proteins to enhance representation and address the issue of limited DTA data.

Unlike previous works~\cite{liu2019roberta} that adopted the widely used pre-training and fine-tuning strategy~\cite{devlin2018bert}, we propose a semi-supervised learning approach that trains on both unlabeled and supervised data. One major limitation of existing methods is that pre-training the molecules and proteins separately does not adequately capture the interaction between drugs and targets, which is critical for DTA prediction. This leads to poor performance when using the pre-training and fine-tuning method for DTA prediction (see Section~\ref{sec:strategies}). 
To address this issue, we propose a semi-supervised multi-task training method that combines regression training on labeled drug-target pairs with masked language modeling (MLM) on both labeled and unlabeled molecules and proteins. 
Similar to paired drug and target sequences, for an unlabeled molecule sequence $M=\{m_i\}_{i=1}^{\vert M \vert}$ and protein sequence $P=\{p_i\}_{i=1}^{\vert P \vert}$, we also take MLM training,
\begin{equation}
    \small
    \mathcal{L}_{mlm}^{M} = - \sum_{s=1}^{S_M} \log P(m_s\|M'), \mathcal{L}_{mlm}^{P} = - \sum_{s=1}^{S_P} \log P(p_s\|P'), \nonumber
\label{eqn:mlm_unlabel}
\end{equation}
where $m_s$, $p_s$ are the masked tokens, $M'$, $P'$ are the masked sequences, $S_M$ and $S_P$ are the number of masked tokens, $log$ is log-likelihood, and $P$ means the conditional probability. We put discussions about our joint training strategy to leverage the unlabeled data and the common pre-training and fine-tuning strategy in Section~\ref{sec:discussion}.

\noindent\textbf{Cost Analysis.} 
Our cross-attention mechanism has an advantage over previous pairwise interaction~\cite{li2020monn} in terms of computational cost. Specifically, our method requires $O((\vert D \vert + \vert T \vert)\times d)$ computations, where $\vert D \vert$ is the drug sequence length, $\vert T \vert$ is the target sequence length, and $d$ is the dimension of the hidden state. The computations for protein-to-drug attention and drug-to-protein attention are $O(\vert D \vert \times d)$ and $O(\vert T \vert \times d)$, respectively. In contrast, previous pairwise attention methods require $O(\vert D \vert \times\vert T \vert \times d)$ computations as each token in one sequence attends to all tokens in the other one, which is more costly. In experiments, we have shown that our cross-attention module is both efficient and effective.

\begin{table*}[t]
  \centering
  \caption{Performance of different methods on BindingDB IC$_{50}$ and $K_i$ datasets. The $\downarrow$ and $\uparrow$ indicate the directions of better results. We report the mean (std) results with three runs.}
  \label{tab:bindingdb_result}
  \scalebox{1.0}{
  \begin{tabular}{lccccc}
      \toprule 
      \textbf{Dataset} & \multicolumn{2}{c}{\bf{IC$\bf{_{50}}$}} & \multicolumn{2}{c}{$\bf{K_i}$} & \\
      \midrule
      \textbf{Method} & \textbf{RMSE$\downarrow$} & \textbf{PC$\uparrow$} & \textbf{RMSE$\downarrow$} & \textbf{PC$\uparrow$} \\
      \midrule 
      Random Forest & 0.910 & 0.780  & 0.970 & 0.780 \\
      DeepAffinity~\cite{karimi2019deepaffinity} & 0.780 & 0.840 & 0.840 & 0.840\\
      DeepDTA~\cite{ozturk2018deepdta} & 0.782 & 0.848 & - & - \\
      MONN~\cite{li2020monn} & 0.764 & 0.858 & - & - \\
      BACPI~\cite{li2022bacpi} & 0.740 & 0.860 & 0.800 & 0.860 \\
      \midrule
      Our baseline & 0.787 (0.002) & 0.848 (0.003) & 0.866 (0.003) & 0.837 (0.002) \\
      SSM-DTA & \textbf{0.712 (0.003)} & \textbf{0.878 (0.004)} & \textbf{0.792 (0.002)} & \textbf{0.863 (0.001)} \\
      \bottomrule 
  \end{tabular}
  }
\end{table*}

\subsection{Training Overview}
For the unlabeled molecule $M$ and protein $P$, the loss functions are $\mathcal{L}_{mlm}^M, \mathcal{L}_{mlm}^P$. Then together with the MLM loss $\mathcal{L}_{mlm}^D, \mathcal{L}_{mlm}^T$ and the MSE loss $\mathcal{L}_{mse}$ on the paired drug-target, the final training objective is
\begin{equation}
    \mathcal{L} = \mathcal{L}_{mse} + \alpha*(\mathcal{L}_{mlm}^D + \mathcal{L}_{mlm}^T) + \beta*(\mathcal{L}_{mlm}^M + \mathcal{L}_{mlm}^P),  \nonumber
\end{equation}
where $\alpha$ and $\beta$ are the coefficients to control the weights of different losses. 
For simplicity, we set $\alpha$ to be the same as $\beta$. 

\subsection{Discussion}
\label{sec:discussion}
The commonly adopted method to leverage unlabeled data is to do pre-train, and the pre-trained model serves as either good initialization (fine-tune-based utilization) or good feature extractors for downstream training (feature-based utilization). Both ways separate the pre-train and the downstream tasks training into two stages, while our strategy is to jointly train the pre-train and the downstream tasks in a multi-task framework and leverage both labeled and unlabeled data together. Two obvious advantages are: (1) Our joint training is super crucial for DTA prediction. The common strategy ignores the interaction during pre-train, which leads to a huge gap between the pre-train and fine-tune prediction~\cite{wang2020bridging} since DTA requires both drug and target for prediction, while the representations learned by our method can well capture the interaction between drug and target. (2) Our method can learn good representations from unlabeled data by MLM training and also the task-specific representations from DTA training, which hence can avoid the catastrophic forgetting problem~\cite{chen2020recall} existed in pre-train/fine-tune based method. 
However, due to the semi-supervised training with large-scale unlabeled data, the training cost is increased compared to directly fine-tuning on an already pre-trained Transformer model (if already existed). From the experimental studies (Section~\ref{sec:strategies}), we find that our method achieves superior performances compared with the two-stage methods. In the future, we will focus on more efficient methods to better leverage the unlabeled data.

\section{Experiments}

\subsection{Experimental Settings}

\textbf{Data.} The unlabeled molecules and proteins are from PubChem~\cite{kim2016pubchem} and Pfam~\cite{mistry2021pfam}. We randomly sampled $10M$ molecules and proteins\footnote{We also study other sizes in Supplementary Materials.} for semi-supervised training. For supervised DTA data, we take from three widely used datasets, BindingDB, DAVIS, and KIBA. BindingDB~\cite{liu2007bindingdb} is a database of measured binding affinities, focusing on the interactions of targets with small drug-like molecules. 
The data is randomly split to 6:1:3 as train/valid/test sets. We study on the IC$_{50}$ and $K_i$ measure as previous works~\cite{karimi2019deepaffinity,li2020monn}. DAVIS~\cite{davis2011comprehensive} contains selectivity assays of the kinase protein family and the relevant inhibitors with their respective dissociation constant ($K_d$) values. KIBA~\cite{tang2014making} dataset includes kinase inhibitor bioactivities measured in $K_i$, $K_d$, and IC$_{50}$ metrics. DAVIS and KIBA are randomly split by 7:1:2 as train/valid/test sets as in DeepPurpose~\cite{huang2020deeppurpose}. 
To further validate the generalization ability of our SSM-DTA, we follow~\cite{lin2022generalizeddta} to evaluate SSM-DTA on DAVIS dataset with unknown drug setting.\footnote{We do not evaluate on KIBA dataset with unknown drug setting as it is not open-sourced by~\cite{lin2022generalizeddta}} The unknown drugs refer to the identification and selection of outliers from all the drug compounds using the substructural features based k-means algorithm~\cite{kim2019pubchem}.
For the affinity label, lower IC$_{50}$/$K_i$/$K_d$ values indicate stronger binding affinities. Following previous works~\cite{karimi2019deepaffinity, huang2020deeppurpose}, to reduce the label range, the concentrations are transformed to logarithm scales: $-\log_{10}(\frac{x}{10^9})$, where $x$ is IC$_{50}$/$K_i$/$K_d$ in the unit of nM. For KIBA dataset, we directly use the KIBA scores provided by the dataset, which is an aggregation of IC$_{50}$, $K_i$, and $K_d$ measurements. Higher KIBA scores indicates stronger binding affinities.
More detailed descriptions about datasets can be found in Appendix.

\noindent\textbf{Model.} The molecule and protein encoders $\mathcal{M_D}$, $\mathcal{M_T}$ are Transformers, and each follows RoBERTa$_{\texttt{base}}$ architecture with $12$ layers. The regression prediction head is a 2-layer MLP with \texttt{tanh} activation function. The compared baseline methods include traditional machine learning methods like linear regression, ridge regression, support vector regression (SVR), KronRLS~\cite{pahikkala2015toward}, and deep learning methods like DeepDTA~\cite{ozturk2018deepdta}, DeepAffinity~\cite{karimi2019deepaffinity}, MONN~\cite{li2020monn}, SAGDTA~\cite{zhang2021sag}, MGraphDTA~\cite{yang2022mgraphdta}, GeneralizedDTA~\cite{lin2022generalizeddta}, BACPI~\cite{li2022bacpi}, GraphDTA~\cite{nguyen2021graphdta}, DeepPurpose~\cite{huang2020deeppurpose} and DeepCDA~\cite{abbasi2020deepcda}.
More details about models and compared baselines models can be found in Appendix.

\noindent\textbf{Evaluation Metrics.} We use (i) mean square error (MSE), (ii) root mean square error (RMSE), (iii) pearson correlation coefficient (PC)~\cite{abbasi2020deepcda}, (iv) concordance index (CI)~\cite{gonen2005concordance}. 
MSE and RMSE measure the difference between ground truth values and values predicted by the model, (v) R-squared ($R^2$)~\cite{lin2022generalizeddta}.
\begin{align}
    MSE(t, p) &= \frac{1}{n}\sum_{i=1}^{n}(t_i - p_i)^2, \nonumber \\
    RMSE(t, p) &= \sqrt{\frac{1}{n}\sum_{i=1}^{n}(t_i - p_i)^2} = \sqrt{MSE(t, p)}, \nonumber
\end{align}
where $t$ and $p$ represent ground truth affinity score and model prediction.
PC measures the linear correlation between ground truth values and values predicted by the model. This measure is varied between -1 and +1. If the two variables are completely correlated it takes +1 and if they are reversely correlated then it takes -1. If there is no correlation between variables then it takes zero~\cite{abbasi2020deepcda}.
\begin{equation}
    R(t, p) = \frac{\sum_{i=1}^{n}(t_i - \bar t)(p_i - \bar p)}{\sqrt{\sum_{i=1}^{n}(t_i - \bar t)^2}\sqrt{\sum_{i=1}^{n}(p_i - \bar p)^2}}, \nonumber
\end{equation}
where $t$ and $p$ represent ground truth affinity score and model prediction.
CI is a model assessment measure introduced by~\cite{gonen2005concordance}. It measures the probability of the concordance between the ground truth value and the predicted value. As suggested in~\cite{pahikkala2015toward}, the CI can be used as an evaluation metric for the prediction accuracy in drug-target interaction task. It can be considered as a generalization of the area under the ROC curve(AUC) that is usually used in binary classification. The intuition behind the CI is whether the predicted binding affinity values of two random drug-target pairs were predicted in the same order as their actual values were or not~\cite{saadat2022drug}.
\begin{equation}
    CI = \frac{1}{Z}\sum_{i,j,i>j}\sigma(t_i > t_j)h(p_i - p_j), \nonumber
\end{equation}
where $t_i$ and $p_i$ denote ground truth affinity score and model prediction of the $i$-th sample, $Z$ denotes the normalization constant that equals the number of data pairs with different label values, and $\sigma$ is a step function that returns one if condition statement is satisfied otherwise it returns zero. Also, $h(x)$ is the Heaviside step function~\cite{pahikkala2015toward} and defined as follows:
\begin{equation}
    h(x) = \begin{cases} 1, & \text {if $x>0$} \\ 
                0.5, & \text{if $x=0$} \\
                0, & \text {if $x<0$}\end{cases}. \nonumber 
\end{equation}
The range of value for CI is between 0 and 1 where the value of one denotes the best result~\cite{abbasi2020deepcda}.
\begin{equation}
    C-index(t, p)=\frac{paired_{correct} + 0.5 \times pairs_{tied}}{pairs_{admissable}}, \nonumber
\end{equation}
where $pairs_{correct}$ is the number of pairs s.t. if $t_x > t_y$, then $p_x > p_y$, pairs; $pairs_{tied}$ is the number of pairs where $p_x = p_y$; and $pairs_{admissable}$ is the number of all possible pairs.The C-index is the average of how often a model says X is greater than Y when, in the observed data, X is indeed greater than Y.
$R^2$ represents the proportion of the variance in the dependent variable that is predictable from the independent variable. The value of $R^2$ ranges from 0 to 1. A higher $R^2$ value indicates a better fit of the model to the data, meaning the model can better explain the variation in the dependent variable.
\begin{equation}
    R^2(t, p) = 1- \frac{\sum_{i=1}^{n}(t_i - p_i)^2}{\sum_{i=1}^{n}(t_i - \bar t)^2}. \nonumber
\end{equation}

Following previous works~\cite{karimi2019deepaffinity, li2020monn, li2022bacpi, lin2022generalizeddta}, results on BindingDB dataset are evaluated on RMSE and PC, results on DAVIS and KIBA datasets are evaluated by MSE and CI scores, and results on DAVIS dataset with unknown drug setting are evaluated on MSE and $R^2$.

\begin{table*}[t]
  \centering
  \caption{Performance of different methods on DAVIS and KIBA datasets. We report the mean (std) results with three runs.}
  \label{tab:davis_kiba_result}
  \scalebox{1.0}{
  \begin{tabular}{lccccc}
      \toprule 
      \textbf{Dataset} & \multicolumn{2}{c}{\textbf{DAVIS}} & \multicolumn{2}{c}{\textbf{KIBA}} & \\
      \midrule
      \textbf{Method} & \textbf{MSE$\downarrow$} & \textbf{CI$\uparrow$} & \textbf{MSE$\downarrow$} & \textbf{CI$\uparrow$} \\
      \midrule 
    Linear Regression & 0.542 (-) & 0.790 (-) & 0.443 (-) & 0.741 (-) \\
      Ridge Regression & 0.532 (-) & 0.800 (-) & 0.443 (-) & 0.742 (-) \\
      SVR & 0.434 (-) & 0.827 (-) & 0.343 (-) & 0.764 (-)\\
      KronRLS~\cite{pahikkala2015toward} & 0.329 (0.019) & 0.847 (0.006) & 0.852 (0.014) & 0.688 (0.003) \\
      GraphDTA~\cite{nguyen2021graphdta} & 0.263 (0.015) & 0.864 (0.007) & 0.183 (0.003) & 0.862 (0.005) \\
      DeepDTA~\cite{ozturk2018deepdta} & 0.262 (0.022) & 0.870 (0.003)  & 0.196 (0.008) & 0.864 (0.002) \\
      DeepPurpose~\cite{huang2020deeppurpose} & 0.242 (0.009) & 0.881 (0.005) & 0.178 (0.002) & 0.872 (0.001) \\
      DeepCDA~\cite{abbasi2020deepcda} & 0.248 (-) & 0.891 (0.003) & 0.176 (-) & 0.889 (0.002) \\
      \midrule
       Our baseline & 0.237 (0.001) & 0.875 (0.003) & 0.162 (0.002) & 0.891 (0.002) \\
      \textbf{SSM-DTA} & \textbf{0.219 (0.001)} & \textbf{0.890 (0.002)} & \textbf{0.154 (0.001)} & \textbf{0.895 (0.001)} \\
      \bottomrule 
  \end{tabular}
  }
\end{table*}

\subsection{Performances on BindingDB}

We first present the results on the BindingDB dataset. 
The IC$_{50}$ and $K_i$ results are shown in Table~\ref{tab:bindingdb_result}. In the table, `Our baseline' refers to our implemented baseline model with two separate encoders, a cross-attention module, and a feed-forward prediction layer, without the semi-supervised multi-task training. From the table, we can see that our SSM-DTA method achieves the best performance in terms of both RMSE and PC metrics. For example, on IC$_{50}$, the RMSE is reduced from $0.787$ to $0.712$ with more than $7\%$ improvement by comparing our baseline and SSM-DTA. When compared with previous state-of-the-art models, such as MONN ($0.764$) and BACPI ($0.740$), our model surpasses their performances by about $3\%-5\%$. On $K_i$, SSM-DTA also achieves the lowest RMSE $0.792$ and highest PC $0.863$. These numbers clearly show the effectiveness of our SSM-DTA framework for binding affinity prediction.

\subsection{Performances on DAVIS and KIBA}
The performance comparison on DAVIS and KIBA datasets is reported in Table~\ref{tab:davis_kiba_result}. 
The sizes of these two datasets are relatively smaller than the BindingDB dataset, but we can also see that our method outperforms previous works with clear improvements. Specifically, on KIBA dataset, the MSE and CI scores of our implemented baseline are $0.162$ and $0.891$, which are already better than most existing works. Our proposed SSM-DTA model further improves the performances to be $0.154$ MSE and $0.895$ CI scores\footnote{We do not directly compare with Affinity2Vec~\cite{thafar2022affinity2vec} and WGNN-DTA~\cite{jiang2022sequence} due to different data splitting method: they use $5$-fold cross-validation and we directly split the dataset proportionally.}. Similar observations are shown on DAVIS dataset. Therefore, we have demonstrated the power of SSM-DTA on both large and small DTA prediction tasks. 

\subsection{Performances on DAVIS with Unknown Drug Setting}
The performance comparison on DAVIS dataset with unknown drug setting is shown in Table~\ref{tab:davis_unk}. Our SSM-DTA model achieves an MSE of 0.8019 and an $R^2$ score of 0.2803, outperforming all baselines. 
Notably, our SSM-DTA shows superior performance than GeneralizedDTA~\cite{lin2022generalizeddta}, which is specially designed for unknown drug settings.
These results indicate that SSM-DTA has superior generalization capabilities on DTA pairs with unknown drugs.

\section{Study and Analysis}
In this section, we present case visualizations of binding activities between drugs and targets, the application of virtual screening and target detection, visualization of drug features, and some other studies to show the impact of our SSM-DTA framework.

\begin{table}[t]
  \centering
  \caption{Performance of different methods on DAVIS dataset with unknown drug setting. * means we select the best model from the original paper.}
  \label{tab:davis_unk}
  \scalebox{1.0}{
  \begin{tabular}{lcc}
      \toprule 
      \textbf{Method} & \textbf{MSE$\downarrow$} & \bf{$R^2\uparrow$} \\
      \midrule 
      DeepDTA~\cite{ozturk2018deepdta} & 1.0271 & 0.1454 \\
      GraphDTA~\cite{nguyen2021graphdta} & 0.8872 & 0.2037 \\
      SAGDTA~\cite{zhang2021sag} & 1.1324 & 0.1654 \\
      MGraphDTA~\cite{yang2022mgraphdta} & 0.8532 & 0.2287 \\
      GeneralizedDTA*~\cite{lin2022generalizeddta} & 0.8467 & 0.2402 \\
      \midrule
      \textbf{SSM-DTA} & \textbf{0.8019} & \textbf{0.2803} \\
      \bottomrule 
  \end{tabular}
  }
\end{table}

\subsection{Case Study of Drug-Target Binding}
Deep learning models often lack interpretability. Based on our cross-attention mechanism, we gain a better understanding of the DTA prediction through interactive attention. In this subsection, we provide two case studies to visualize the atomic level attention on compound molecules and the amino acid level attention on proteins. 
For atomic level attention, we carry out the experiment to see the attention values of target$\to$drug attention calculation, where the query is target protein and the key/value is compound. Therefore, the attention weights on each atom reflect the importance of the compound atoms for one specific target. Similarly, for amino acid level attention, the experiment is to see attention values of drug$\to$target attention, where the query is compound and the key/value is target protein. The attention weights on each amino acid of a protein can reflect the importance of these amino acids to the corresponding drug compound. 

\begin{figure*}
     \centering
     \begin{subfigure}[t]{0.45\linewidth}
         \includegraphics[width=\linewidth]{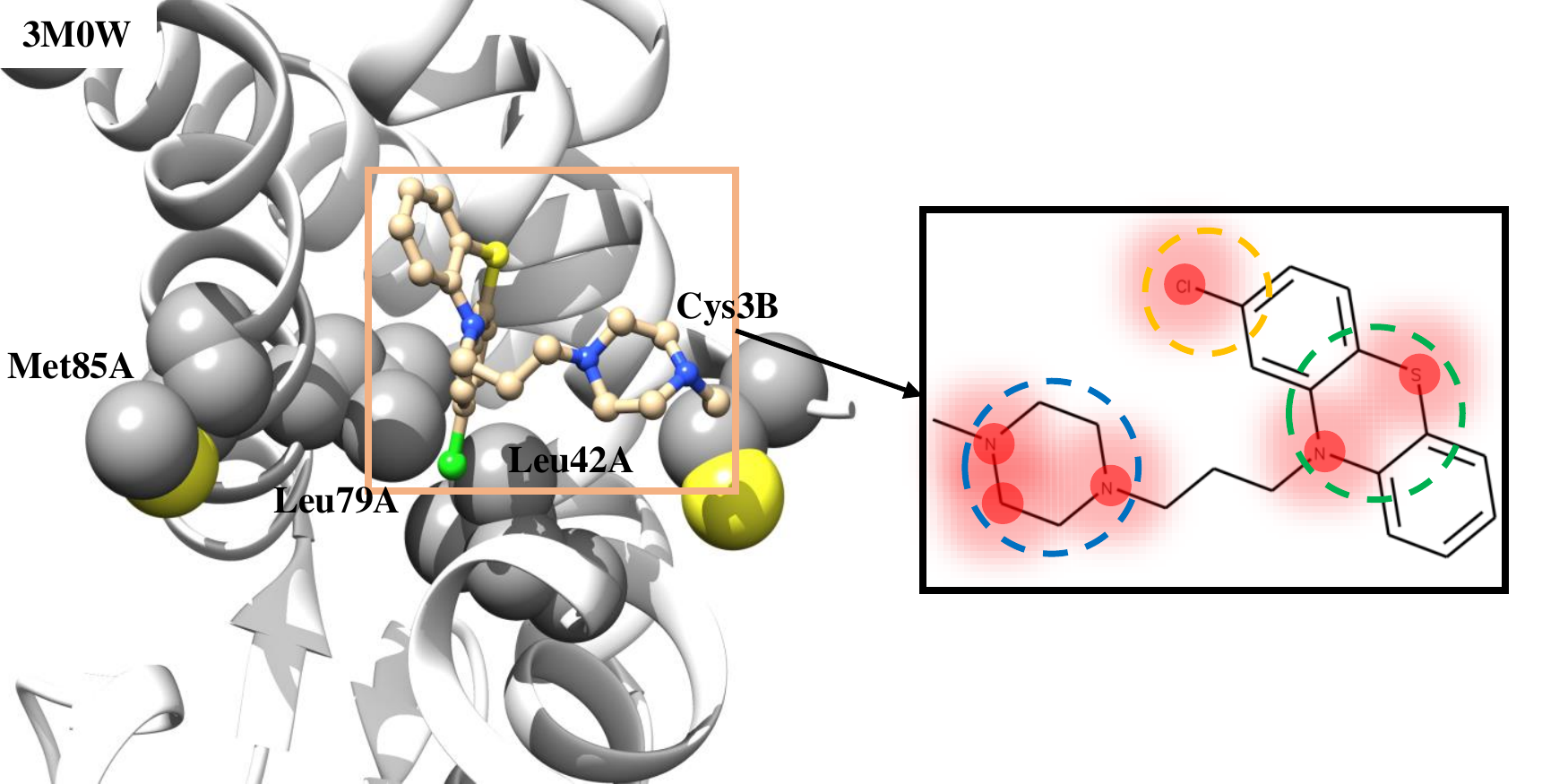}
         \caption{S100A4 target and prochlorperazine (PCP). The left panel shows the 3D view of S100A4 with PCP. The right panel shows the 2D view of PCP, whose atoms with high attention scores are highlighted. The pharmacophore groups according to SAR are in dashed circles: (1) The blue circle is a nitrogen-containing basic group. The side chain substituted with piperazine has the strongest effect; 
         (2) The green circle contains sulfur at 5-position and nitrogen at 10-position, which are associated with antipsychotic activity; (3) The yellow circle is an electron withdrawing group at 2-position enhancing drug activity.}
         \label{fig:case_3M0W}
     \end{subfigure} \qquad
    \begin{subfigure}[t]{0.45\linewidth}
         \includegraphics[width=\linewidth]{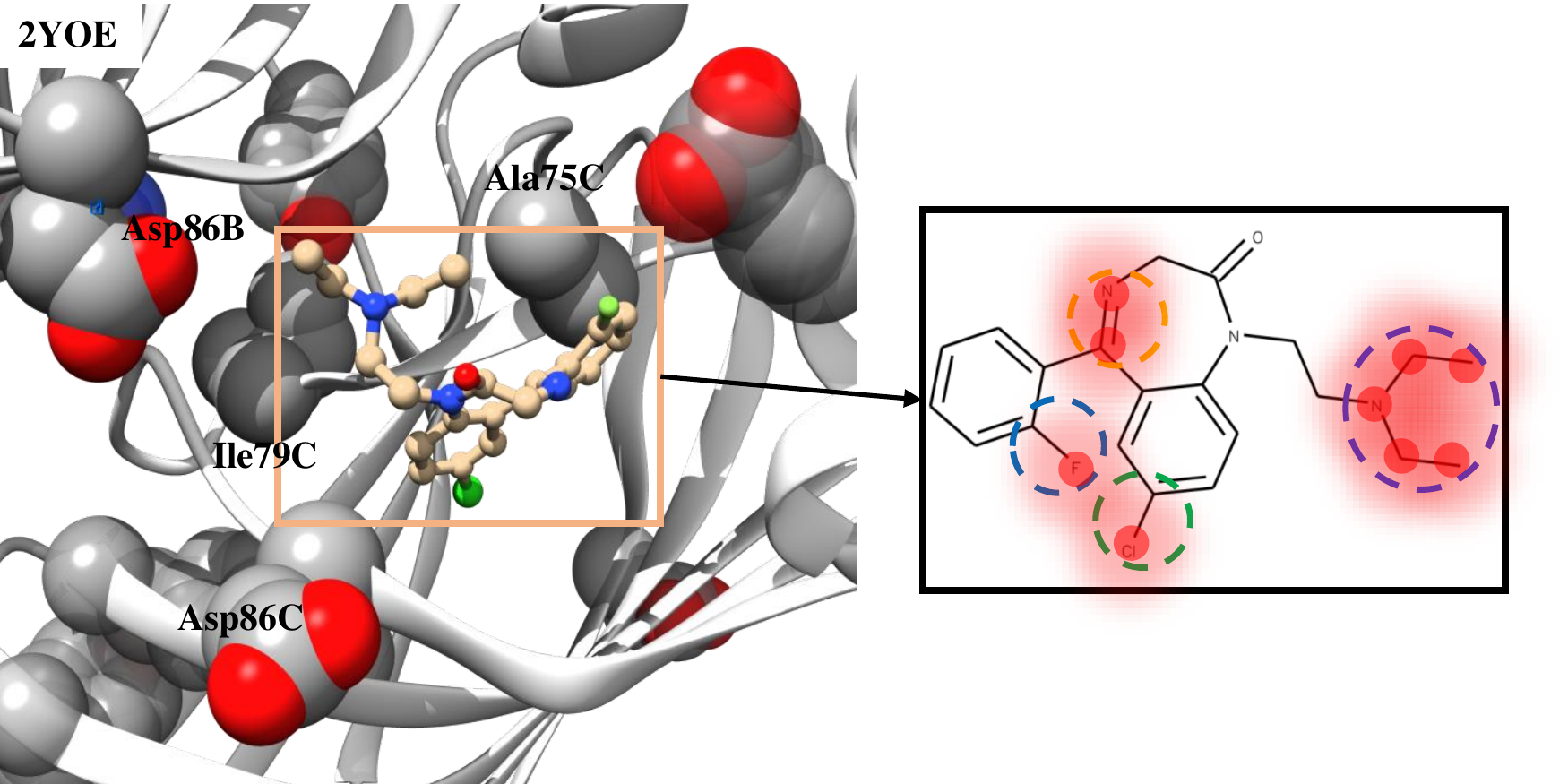}
         \caption{GABA$_A$ receptor and flurazepam. The 3D structure of the complex (PDB ID: 2YOE) and 2D representation of the flurazepam are shown on the left and right panels. Atoms of flurazepam with high attention scores are highlighted in red color, and the important groups according to SAR are circled: (1) The blue and green circle are electron-withdrawing groups that enhance activity; (2)The yellow circle is a saturated double bond at 4,5-position, which increases sedative and antidepressant effects; and (3) The group enclosed by the purple circle prolongs the efficacy.}
         \label{fig:case_2YOE}
     \end{subfigure}
 \caption{Case studies of protein targets and the corresponding ligands. The amino acids with high attention scores are shown in van der Waals representations and the drug compound molecules are in ball-stick format in 3D models. The amino acids near the binding sites are labeled with their names and residue indexes.}
\end{figure*}

Following~\cite{chen2020transformercpi}, we choose the drug prochlorperazine (PCP) and its target S100A4 (UniProt ID), whose atomic structure is experimentally determined (PDB ID: 3M0W)\footnote{\url{https://www.rcsb.org/structure/3M0W}}, for one example analysis. Prochlorperazine is a phenothiazine antipsychotic medicine used to treat anxiety or schizophrenia, and its structure-activity relationship (SAR) has been thoroughly explored. From Fig.~\ref{fig:case_3M0W}, we can see that the attention highlighted atoms of PCP are consistent with the SAR features, demonstrating that our model is capable of capturing key atomic groups interacting with proteins. The ground truth of structural binding site information can be clearly visualized from the protein-drug complex structure (PDB ID: 3M0W) in the same figure.
Among 10 residues (out of 100) with the highest attention scores, 4 are located in the vicinity of the binding site.  Interestingly, 3 residues (Leu42, Leu79, and Met85) are from one chain, and one residue (Cys3) is from the other chain.
The second case is the GABA$_A$ receptor, a ligand-gated ion channel from Erwinia chrysanthemi (ELIC) (UniProt ID: P0C7B7). We analyze the interaction between GABA$_A$ and the drug flurazepam, whose key atoms are highlighted in Fig.~\ref{fig:case_2YOE}, corresponding to four atomic groups. Similar to the first case, the prediction results are assessed using the complex structure determined with crystallography method (PDB ID: 2YOE)\footnote{\url{https://www.rcsb.org/structure/2YOE}}. As shown in Fig.~\ref{fig:case_2YOE}, four of the amino acids around the binding sites are identified based on the attention score using the flurazepam as query input. These identified amino acids are labeled to emphasize their close contact with the drug molecule. 

These case studies clearly demonstrate the power of our model in identifying important atomic groups or amino acids. We would like to stress that the structure information was only used when assessing the predicted key amino acids in these case studies instead of our training method. Although other amino acids not within the immediate vicinity of the binding site may also have higher attention scores, we found that the model prediction results are significantly meaningful (see the above S100A4-PCP interactions). These results suggest that the SSM-DTA model learns important rules in drug-target interactions via the proposed training framework.

\subsection{Ablation Study}
We conduct an ablation study to verify the effectiveness of each component: the MLM multi-task training, the cross-attention module, and the semi-supervised training with unlabeled data. The results are presented in Table~\ref{ablation} and show the valid and test MSE/RMSE and Pearson Correlation scores on the BindingDB IC$_{50}$ dataset. In the table, `Paired MLM' refers to the MLM training performed only on paired drug-target data, and the cross-attention module is the default module in our baseline model. From the table, we observe the following: (1) MLM multi-task training on the original paired data plays an important role in DTA prediction. For instance, the valid MSE is decreased from $0.625$ to $0.545$ with `Paired MLM', and the test RMSE $0.735$ is already the best performance among previous works, e.g., BACPI ($0.740$). (2) Semi-supervised training on unlabeled data (without `Paired MLM') also improves the performance, e.g., valid MSE reduction from $0.625$ to $0.601$ and test RMSE from $0.787$ to $0.772$. (3) With all components, the valid/test results are further improved to $0.513$ and $0.712$. Similar observations are shown on the PC metric. Therefore, it is obvious that each component has a positive effect on improving DTA prediction.

\begin{table}[t]
\begin{center}
\caption{Ablation study of SSM-DTA. `PC' is Pearson Correlation. }
\label{ablation}
\resizebox{\linewidth}{!}{
\begin{tabular}{c c c c c}
    \toprule%
    \multicolumn{3}{c}{\textbf{Module}} & \textbf{MSE/RMSE$\downarrow$} & \textbf{PC$\uparrow$} \\
    \midrule
      Cross-attention  & Paired MLM & Unlabeled Data & Valid/Test &  Valid/Test  \\
      \midrule 
      \cmark & \xmark & \xmark & 0.625/0.787 & 0.846/0.848  \\
      \cmark & \xmark & \cmark & 0.601/0.772 & 0.853/0.855 \\
      \cmark & \cmark & \xmark & 0.545/0.735 & 0.866/0.868 \\
      \cmark & \cmark & \cmark & \textbf{0.513/0.712} & \textbf{0.875/0.878} \\
      \bottomrule 
\end{tabular}
}
\end{center}
\end{table}

\subsection{Training Strategies Comparison}
\label{sec:strategies}
As discussed, we implemented a different strategy to leverage the large-scale unlabeled data. This is different from the common pre-train (on large-scale unlabeled data) and fine-tune (on supervised labeled data) strategies (Section~\ref{sec:discussion}). To compare these different strategies, we conduct a study on the BindingDB IC$_{50}$ dataset. The following three different training strategies are compared: (1) \textit{Feature-based tuning}. Pre-training on large-scale unlabeled data, and the pre-trained model is fixed and used as a feature extractor to conduct subsequent tuning; (2) \textit{Pre-training and fine-tuning}. Pre-training on large-scale unlabeled data, the pre-trained model and newly added cross-attention are trained to fine-tune DTA prediction; and (3) \textit{Semi-supervised multi-task training}. This is our SSM-DTA strategy. We present the results in Table~\ref{diff_train_method} for both valid and test datasets. The feature-based training strategy does not perform well due to the limited number of tuning parameters (only the newly added cross-attention module was fine-tuned). Our SSM performs best among the three strategies, outperforming the general pre-training and fine-tuning method. This supports our claim that separate pre-training on molecules or proteins ignores the importance of interaction for DTA prediction, and highlights our SSM is a better choice for paired interaction-related tasks than pre-training and fine-tuning strategy.

\begin{figure*}[!h]
     \centering
     \begin{subfigure}[b]{0.32\textwidth}
         \centering
         \includegraphics[width=\textwidth]{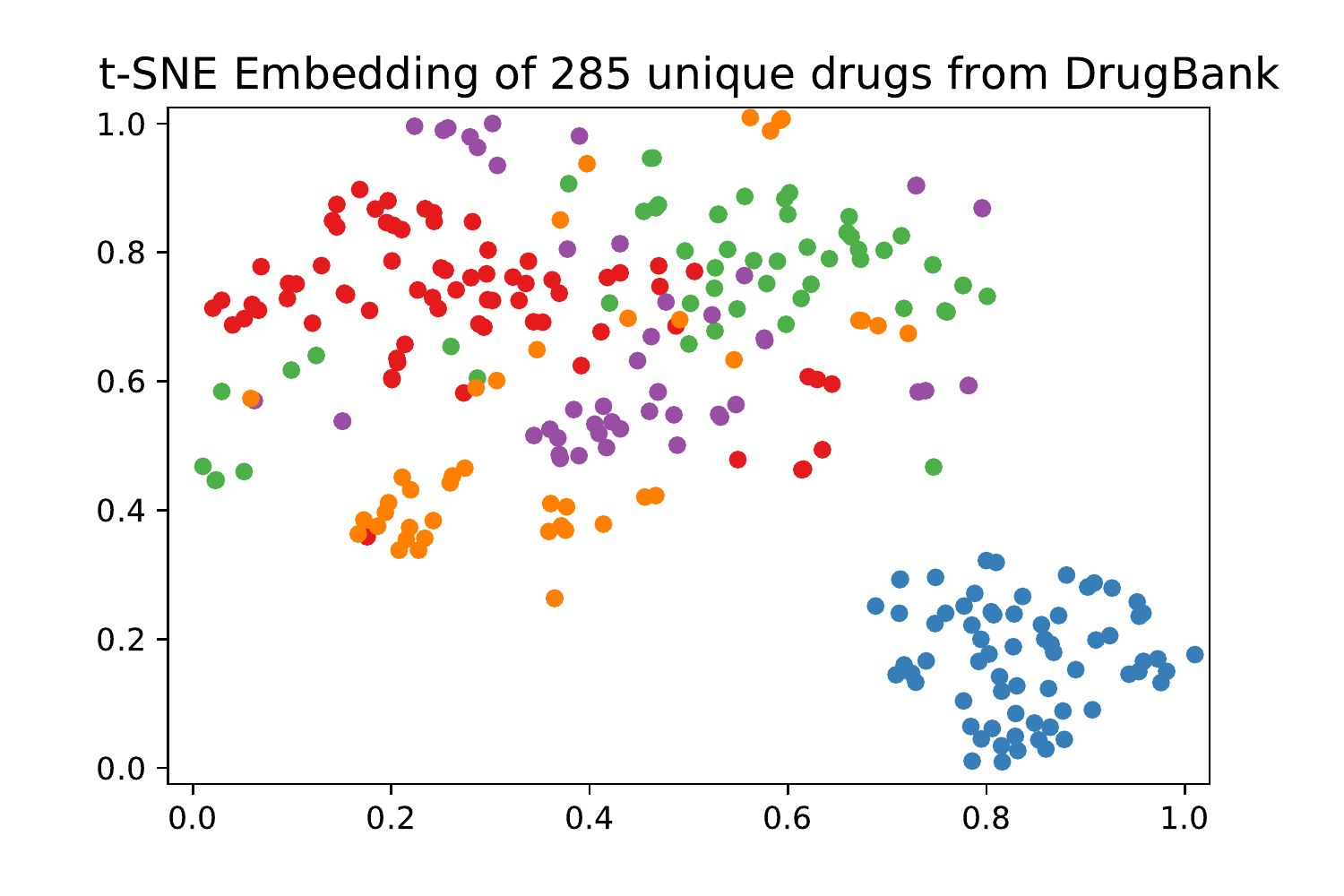}
         \caption{Embeddings from pre-trained DVMP.}
         \label{fig:dvmp_drug}
     \end{subfigure}
     \begin{subfigure}[b]{0.32\textwidth}
         \centering
         \includegraphics[width=\textwidth]{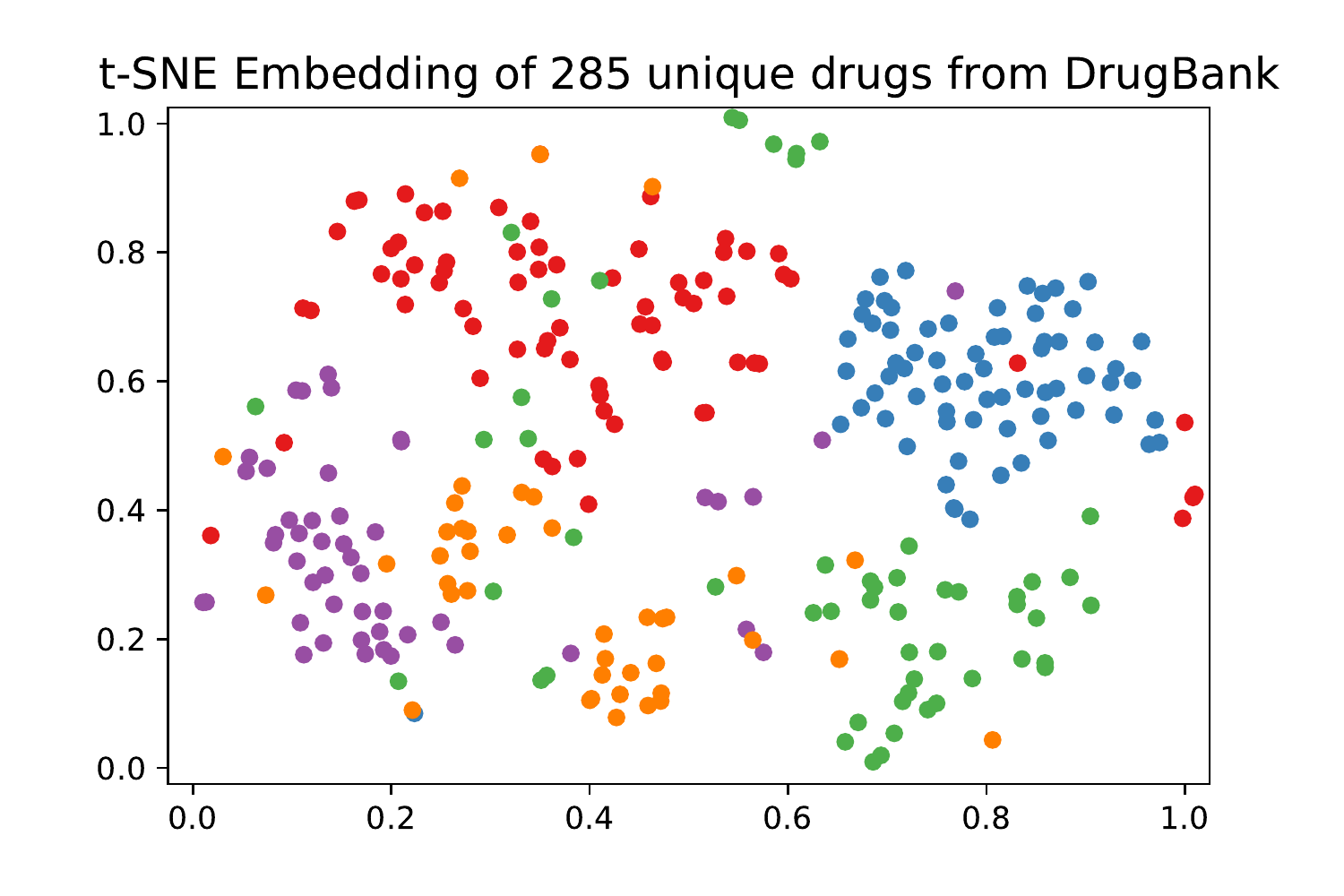}
         \caption{Embeddings from our baseline model.}
         \label{fig:base_drug}
     \end{subfigure}
     \begin{subfigure}[b]{0.32\textwidth}
         \centering
         \includegraphics[width=\textwidth]{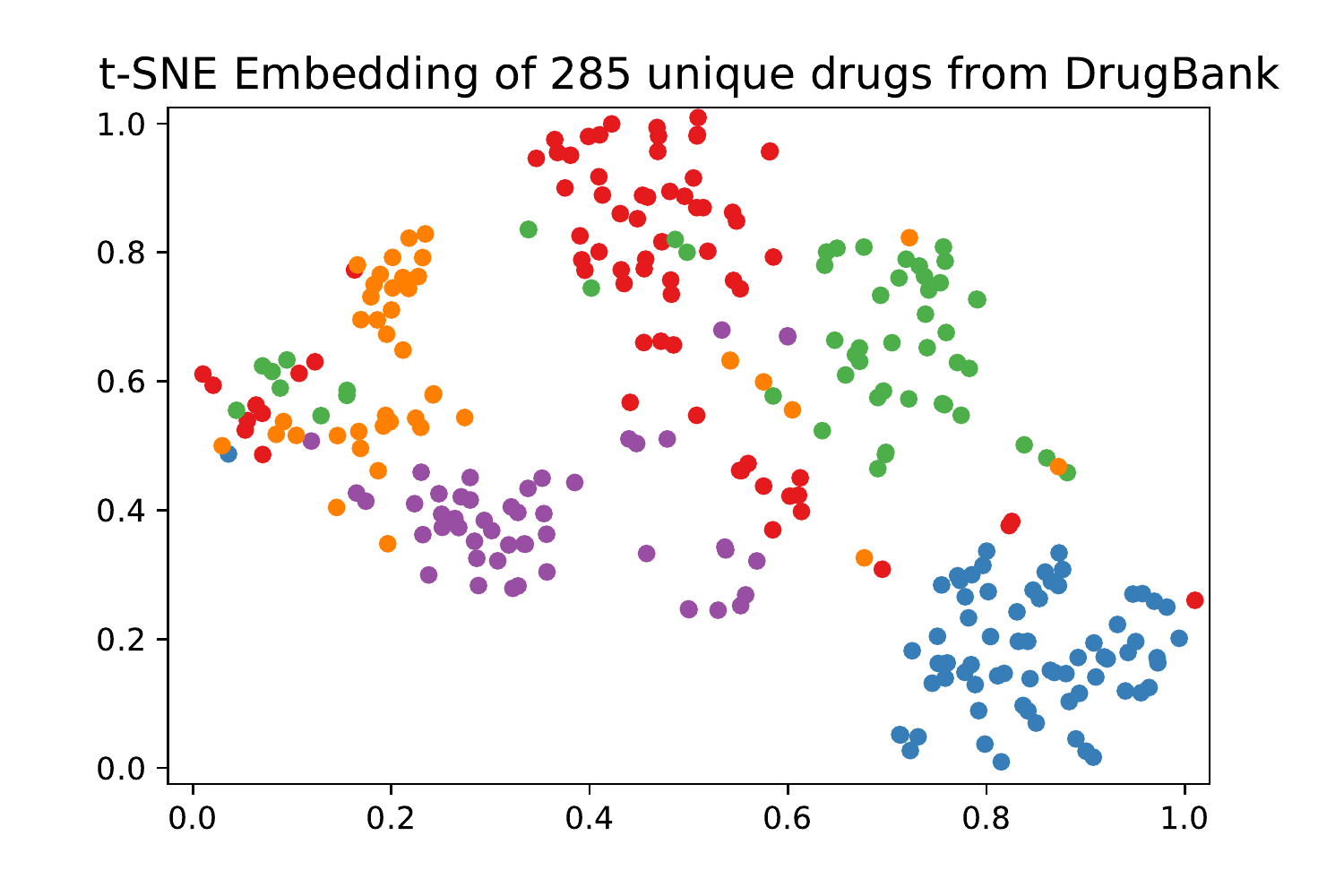}
         \caption{Embeddings from our SSM-DTA model.}
         \label{fig:SSM_drug}
     \end{subfigure}
     \caption{Visualized embeddings of $285$ drugs that correspond to $5$ targets from DrugBank dataset. Each color represents the drugs for one specific target. (a) shows the drug embeddings from the DVMP pre-trained model, (b) is from our baseline model, and (c) is from our SSM-DTA. }
     \label{fig:embedding}
\end{figure*}

\begin{table}[h]
  \centering
  \caption{Different training strategies to leverage the unlabeled molecule and protein data. `PC' stands for Pearson Correlation.}
  \label{diff_train_method}
  \resizebox{\linewidth}{!}{
  \begin{tabular}{lccccc}
        \toprule
        \textbf{Training Strategies} & \textbf{Valid MSE$\downarrow$} & \textbf{Valid PC $\uparrow$} & \textbf{Test RMSE$\downarrow$} & \textbf{Test PC $\uparrow$} \\
        \midrule
        Our baseline & 0.545 & 0.866 & 0.735 & 0.868 \\
        \midrule
        Feature-based tuning & 0.638 & 0.840 & 0.795 & 0.843 \\
        Pre-training/fine-tuning & 0.536 & 0.868 & 0.738 & 0.867 \\
        SSM-DTA & \textbf{0.513} & \textbf{0.875} & \textbf{0.712} & \textbf{0.878} \\
        \bottomrule
  \end{tabular}
  }
\end{table}

\subsection{Target Detection on DrugBank}
To evaluate the application values and generalization capability of our model, we perform an experiment using the DrugBank~\cite{wishart2006drugbank} dataset, where most drugs and targets are excluded from the BindingDB training data. DrugBank contains the real-world drug-target pairs with different interaction types, and we take the drug-target pairs with `inhibitor' like\footnote{Including `aggregation inhibitor', `weak inhibitor', `inhibitory allosteric modulator', `inhibitor' and `translocation inhibitor'.} interactions, whose activity is quantified as IC$_{50}$ scores. 
The DrugBank dataset includes $4351$ targets, which interact with at least one drug. We carry out the test by limiting the drug dataset to those that interact with less than 20 target proteins (most of the drugs only interact with one target, making it more difficult to identify true targets). We randomly selected $100$ drugs among those that satisfy the selection criteria, and predict the affinity scores between $100$ drugs and $4351$ targets.
For each drug, the targets are ranked based on the predicted affinity scores, and we analyze the ranking position of the true targets out of $4351$ candidates. The performance is evaluated by counting the number of correctly identified targets for these $100$ drugs in the top-ranked candidates (labeled as Top-$N$).
For these $100$ drugs in this test, $\bf{13}$ drugs have correctly identified their best candidate target (i.e., Top-$1$), a significant improvement compared to the baseline model that only predicts $\bf{5}$ pairs in the Top-$1$ category. If the candidate pool is relaxed to Top-$5$, our SSM-DTA model correctly identifies the targets for $\bf{18}$ drugs, while the baseline model finds $\bf{10}$. This test result shows that our SSM-DTA model can be potentially useful in drug repurposing research, by predicting the targets for a given drug.

\subsection{Virtual Screening on EGFR}
A practical scenario for DTA prediction is virtual screening, which involves screening a large number of molecules to identify potential drugs for a specific target. This is the opposite application of target detection discussed in the above section. In this study, we conduct an experiment to evaluate the screening ability of our DTA model. Specifically, we choose the Epidermal Growth Factor Receptor (EGFR) (PDB ID: P00533)\footnote{\url{https://www.uniprot.org/uniprotkb/P00533/entry}} as the study target, since it is an important gene target that highly related to the cell lung cancer. For all the drugs in the DrugBank (total number $6848$), we calculate the binding affinity between each drug and EGFR target sequence, then we rank all the drugs based on the predicted affinity score. From the top-ranked drugs, we find that Afatinib (DrugBank ID: DB08916)\footnote{\url{https://go.drugbank.com/drugs/DB08916}} is ranked $\bf{8^{th}}$ among all the $6848$ drugs (top $0.1\%$), and \emph{`Afatinib is an antineoplastic agent used for the treatment of locally advanced or metastatic non-small cell lung cancer (NSCLC) with non-resistant EGFR mutations or resistance to platinum-based chemotherapy'} (from DrugBank introduction). This high ranking clearly demonstrates that our DTA prediction has great potential for virtual screening.

\subsection{Drug Feature Learning and Drug Grouping}
In reality, multiple drugs can interact with the same target, and these drugs often bind to similar regions on the target. This intrinsic correspondence means that drugs for the same target possess some common properties that may not be directly observable with conventional statistical parameters. Deep learning models, however, are capable of capturing hidden features, which are more efficient in describing drug properties. To evaluate this, we grouped $285$ drugs that interact with $5$ targets from the DrugBank dataset based on the learned features of three models: our baseline model, the SSM-DTA trained model, and a strong pre-trained molecule model (DVMP~\cite{zhu2021dual}). The embedding of these drugs in the manifold is shown in Fig.~\ref{fig:embedding}, with colors indicating the corresponding targets.
From the figure, we can see that both the baseline DTA and SSM-DTA models show distinguishable clusters, which can be mapped to the corresponding targets. In the case of the pre-trained DVMP model, only the two apparent clusters can be identified (the blue dots and the rest), suggesting that the target-specific features are not learned by the model. The SSM-DTA model further improves drug embedding, manifested in better-defined clusters than those of the baseline model. For example, the drugs in the green/orange groups are broadly spread in the baseline model embedding, but they are centralized into clusters in the SSM-DTA model embedding (Fig.~\ref{fig:SSM_drug}). This test result demonstrates that the drugs can be better grouped according to their targets. Such target-specific features learned by the SSM-DTA model can be a foundation for improved performance in DTA prediction.

\section{Deeper Understanding of SSM-DTA}
\subsection{Training Analysis}
\begin{figure}[!ht]
\centering
    {\includegraphics[width=0.95\linewidth]{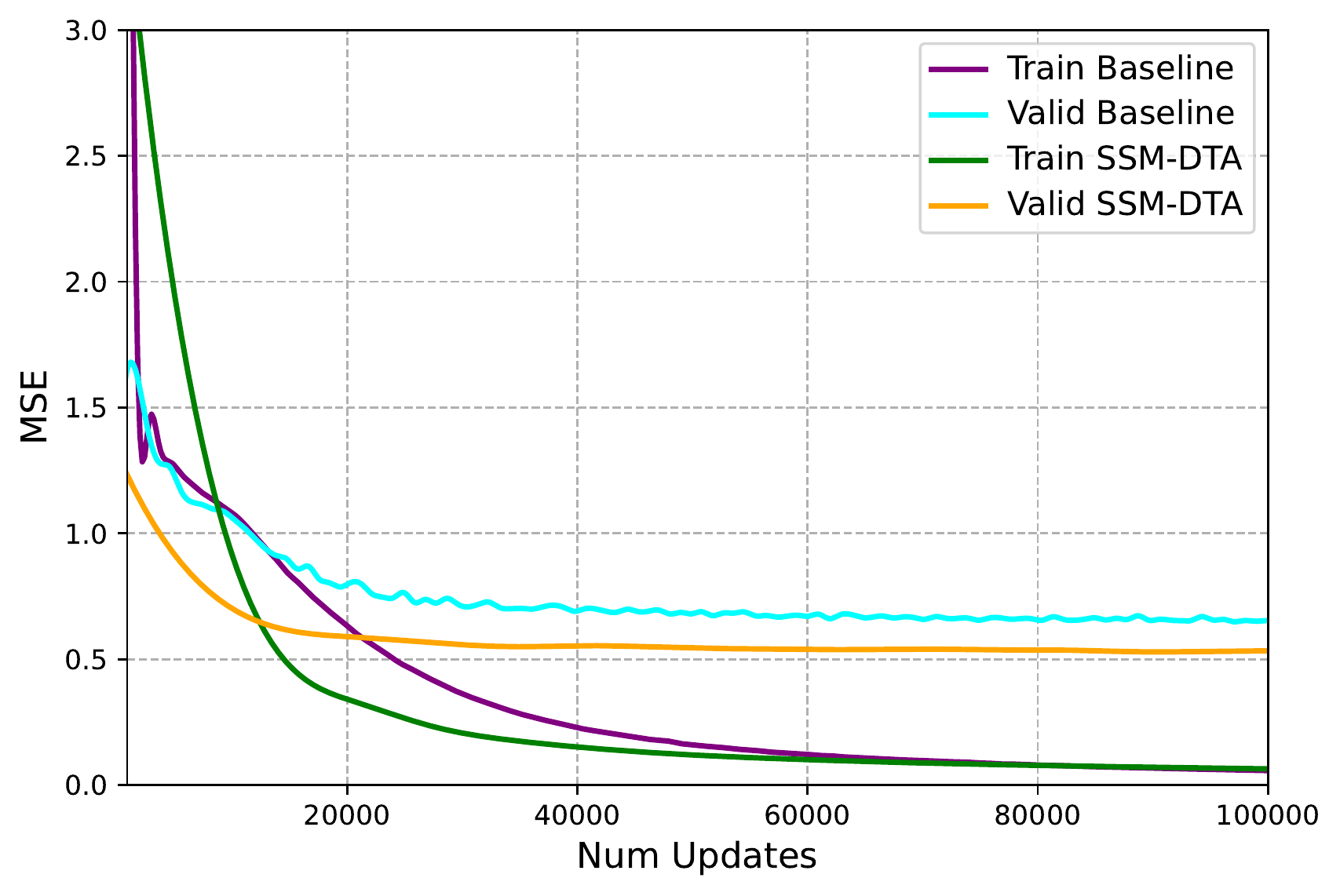}}
    \caption{The MSE loss curves on BindingDB IC$_{50}$ training and validation datasets along the training process. Results are from our implemented baseline model and our SSM-DTA method.}
    \label{fig:loss_curve}
    \vspace{-0.5cm}
\end{figure}

We provide the training process analysis from the MSE value on both training and validation sets. We plot the corresponding values at each iteration in Fig~\ref{fig:loss_curve} for both our baseline and SSM-DTA models. From the curve, we can observe that along the training process, the MSE values on both training and validation sets from our SSM-DTA method are lower than the baseline model. While the converged training MSE values are similar on the training set, the validation MSE of SSM-DTA model is much better than the baseline model. Besides, the SSM-DTA model converges faster than the baseline model. 
These curves demonstrate that model training can benefit from the proposed SSM-DTA framework to get faster convergence and higher accuracy and to achieve better performance in validation and test datasets. 

\begin{figure}[htbp]
     \centering
     \begin{subfigure}[b]{0.38\textwidth}
         \centering
         \includegraphics[width=\textwidth]{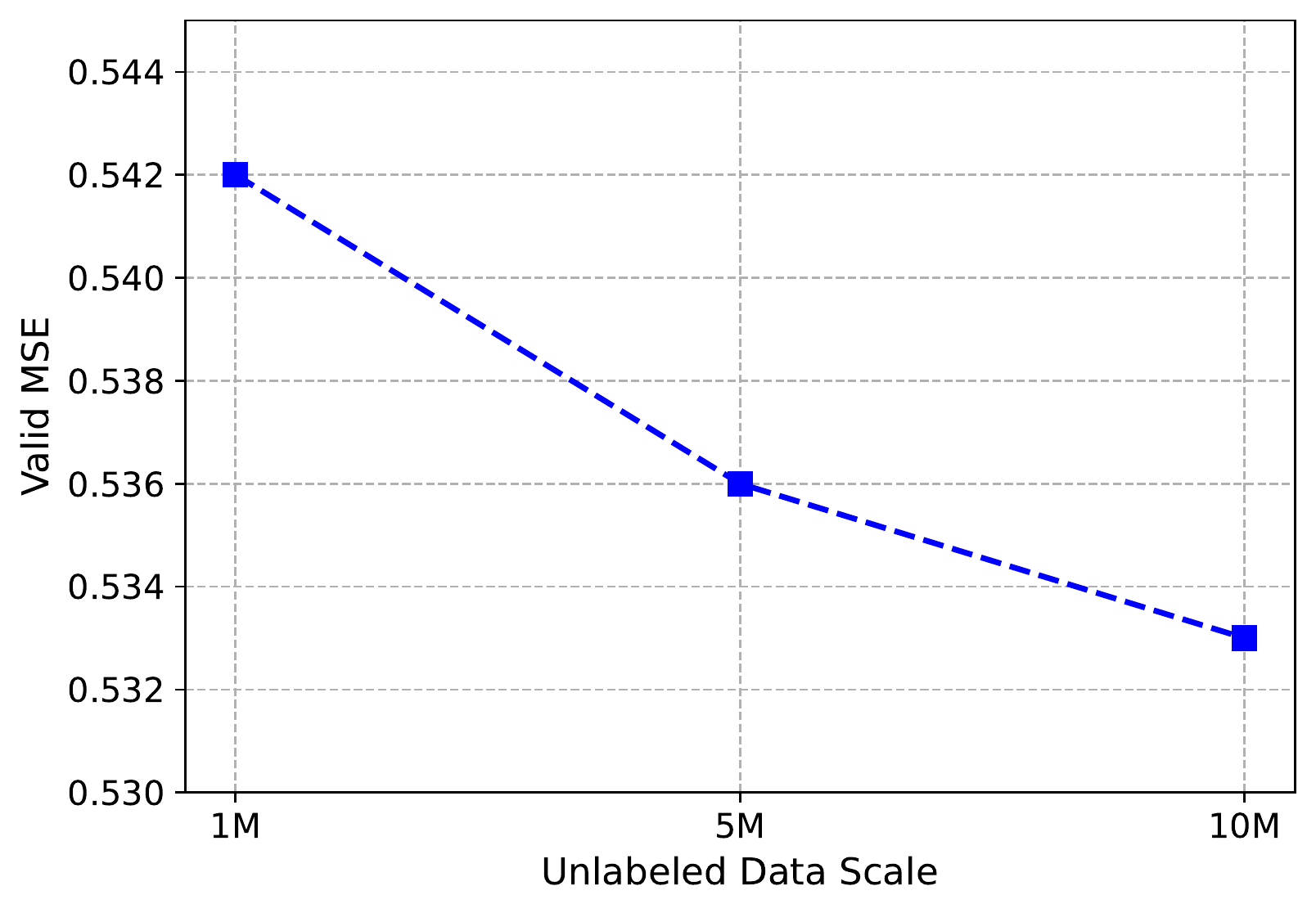}
         \caption{Varied scales of unlabeled protein and molecule.}
         \label{fig:unlabeled_data_scale}
     \end{subfigure}
     \begin{subfigure}[b]{0.38\textwidth}
         \centering
         \includegraphics[width=\textwidth]{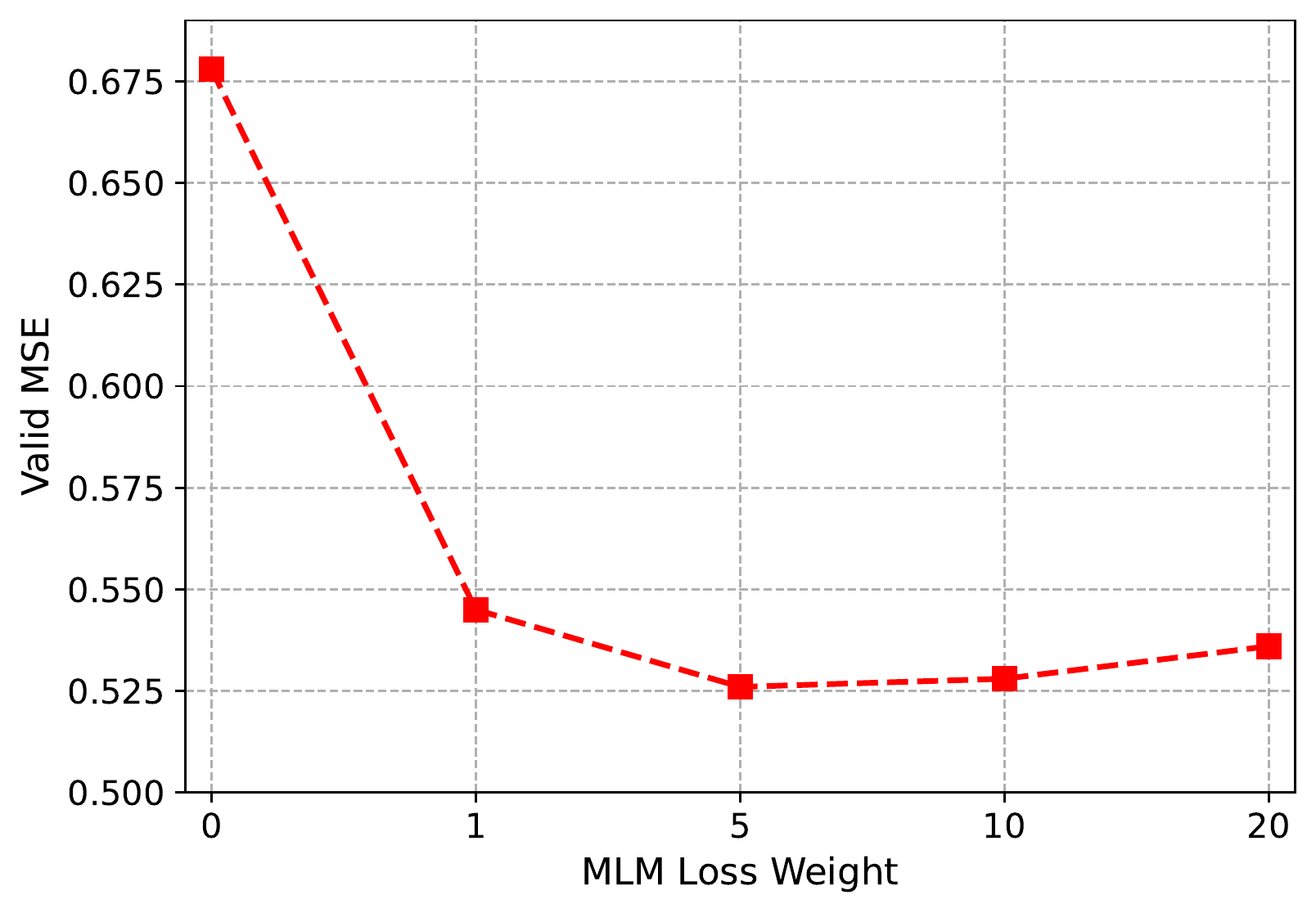}
         \caption{Varied coefficient weight $\alpha$ and $\beta$, here $\alpha=\beta$.}
         \label{fig:diff_mlm_w}
     \end{subfigure}
     \caption{Effects of (a) varied scales of unlabeled molecule and protein data, and (b) varied MLM loss coefficient $\alpha$ and $\beta$.}
     \label{fig:diff_data_scale_mlm_w}
     \vspace{-0.5cm}
\end{figure}

\subsection{Effects of Unlabeled Data and MLM Loss Weight}
We first study the effects of the different data scales of the unlabeled molecules and proteins. Specifically, we vary the unlabeled dataset to be $1M$, $5M$, and $10M$ scales during training and evaluate the effects of these unlabeled data scales on the performance of the trained model. The validation MSE scores are reported in Fig. \ref{fig:unlabeled_data_scale}. The figure shows that a larger unlabeled data scale will gradually improve the performance. Due to the computation resource limitation, we do not perform experiments on larger datasets. We suspect the reason for increased performance comes from data diversity. The larger unlabeled data can help the model to learn better-generalized representations from more diverse data so as to enhance the DTA prediction task.

Our method introduces the coefficient weights $\alpha$ and $\beta$ for MLM training, and we simply set $\alpha$ to be the same as $\beta$. Therefore, it is also necessary to investigate its effect. Here the weight value is varied among $\{0, 1, 5, 10, 20\}$.
Note that we do not use the cross-attention module here. Fig. \ref{fig:diff_mlm_w} shows the results of validation MSE under different settings. We can see that the weight has a trade-off effect and the optimal weight is $5$, around which the MSE value for the validation set is the lowest. This is expected since the goal of MLM training is different from DTA prediction and the reduction in the DTA prediction MSE requires a proper incorporation with MLM training. 

\subsection{Effectiveness of our Cross-attention Module}
To show the effectiveness of our proposed cross-attention module, we replace it with pairwise attention that was used in DeepAffinity~\cite{karimi2019deepaffinity} for an experiment, and the results are shown in Table~\ref{tab:diff_interaction}. The better results of SSM-DTA show that our cross-attention is not only more effective (better performance) but also more efficient (lower computation cost as discussed before).

\begin{table}[t]
  \centering
  \caption{Performance comparison on DAVIS and BindingDB K$_i$.}
  \label{tab:diff_interaction}
  \resizebox{\linewidth}{!}{
  \begin{tabular}{lccccc}
      \toprule 
      \textbf{Dataset} & \multicolumn{2}{c}{\textbf{DAVIS}} & \multicolumn{2}{c}{\textbf{BindingDB K$_i$}} & \\
      \midrule
      \textbf{Method} & \textbf{MSE$\downarrow$} & \textbf{CI$\uparrow$} & \textbf{RMSE$\downarrow$} & \textbf{PC$\uparrow$} \\
      \midrule 
       Pairwise Attention & 0.233 (0.002) & 0.884 (0.002) & 0.802 (0.001) & 0.852 (0.002) \\
      \textbf{SSM-DTA} & \textbf{0.219 (0.001)} & \textbf{0.890 (0.002)} & \textbf{0.792 (0.002)} & \textbf{0.863 (0.001)} \\
      \bottomrule 
  \end{tabular}
  }
\end{table}

\section{Conclusion}
Predicting Drug-Target Affinity (DTA) is crucial in the drug discovery process, but it has been difficult to achieve using deep learning approaches due to limitations in supervised data. To address this issue, we propose three strategies to improve DTA prediction performance. We develop an SSM-DTA model that combines paired MLM training with a semi-supervised multi-task framework that leverages large-scale unlabeled data and an efficient cross-attention module for drug-target interaction. Experiments on multiple DTA benchmark datasets show improved performance of our method. We also demonstrate the ability of our model to generalize to unknown drugs and identify key atom groups and amino acids through three case studies. Additionally, we explore and demonstrate the target-specific features embedded in the SSM-DTA by grouping drugs according to their targets, providing an explanation for the improved performance of our proposed method. In the future, we would like to explore the unification of structure and sequence methods in an efficient way. 

\section{Limitations}
Though our SSM-DTA framework achieves superior results on DTA tasks, it still has several limitations.
First, as SSM-DTA is trained on both labeled and large-scale unlabeled data at the same time, the training cost is relatively high. For example, for BindingDB IC50 (training set size: 263,583), the overall training process was done on 8 NVIDIA V100 GPUs for 4.8 days.
Second, while SSM-DTA performs well on benchmark datasets, actual validation of this method would require experimental results. Without such validation, there's a risk of overfitting the specific datasets and not truly predicting drug-target affinity in real-world applications.
Moreover, a predominant challenge with most drug-target affinity (DTA) prediction methods is their inability to generalize across different datasets, primarily due to the diversity in drugs and targets and different affinity types present within these datasets.
These limitations are important for future exploration.

\section{Author Contributions Statement}
Q.P. and L.W. conceived the experiments, Q.P. conducted the experiments, Q.P., L.W., and J.Z. analysed the results. Q.P. and L.W. wrote the manuscript, Y.X., S.X., and R.Y. reviewed the manuscript.

\section{Acknowledgments}
The authors thank the anonymous reviewers for their valuable suggestions. 

\section{Data Availability}
The code and data for this study are available at \url{https://github.com/QizhiPei/SSM-DTA}.

\section{Funding}
This work was partially supported by National Natural Science Foundation of China (NSFC Grant No. 62122089), Beijing Outstanding Young Scientist Program NO. BJJWZYJH012019100020098, and Intelligent Social Governance Platform, Major Innovation \& Planning Interdisciplinary Platform for the ``Double-First Class'' Initiative, Renmin University of China, the Fundamental Research Funds for the Central Universities, and the Research Funds of Renmin University of China. 

\bibliographystyle{unsrt}
\bibliography{reference}

\clearpage

\begin{appendices}

\section{Experimental Settings}

The detailed experimental settings of our binding affinity prediction are introduced, including the datasets, model training, and compared baseline models.
\subsection{Datasets}
Our method leverages both the limited DTA data and the large-scale unlabeled molecule and protein data. The introductions are as follows. \\
\newline
\noindent{\bf{Drug-Target Affinity Prediction Datasets}}
We use the widely adopted three benchmark datasets for DTA prediction, which are BindingDB, DAVIS, and KIBA datasets.  \\
\newline
\noindent{\bf BindingDB}~\cite{liu2007bindingdb} is a public database\footnote{\url{https://www.bindingdb.org/}} of measured binding affinities, focusing on the interactions of targets with small drug-like molecules. Previous works like DeepAffinity~\cite{karimi2019deepaffinity}, MONN~\cite{li2020monn} and BACPI~\cite{li2022bacpi} have been evaluated on half-maximal inhibitory concentration (IC$_{50}$) values derived from the BindingDB database. For consistency purposes, we use the same BindingDB dataset in the test. The dataset contains 376,751 IC$_{50}$-labeled samples, with 255,328 unique drugs and 2,782 unique targets. We randomly sample 70\% for training (including 10\% held out for validation), and 30\% for test. In addition, there are three other measurements: concentration for 50\% of maximal effect (EC$_{50}$), inhibition constant ($K_i$) and dissociation constant ($K_d$), where the data size is smaller than IC$_{50}$. In our experiments, we also study the $K_i$ measurement. The same data splitting and labeling protocols are applied.
To reduce the label range, the concentrations are transformed to logarithm scales: $-\log_{10}(\frac{x}{10^9})$, where $x$ is IC$_{50}$ or $K_i$ in the unit of nM.

\noindent{\bf DAVIS}~\cite{davis2011comprehensive} dataset contains selectivity assays of the kinase protein family and the relevant inhibitors with their respective dissociation constant ($K_d$) values. It has 30,056 DTA pairs with 68 unique drugs and 379 unique targets. Following DeepPurpose~\cite{huang2020deeppurpose}, we randomly sample 70\% as the training set, 10\% as the validation set, and 20\% as the test set. Same as~\cite{he2017simboost,ozturk2018deepdta}, we also transform the raw values to logarithm scales in the same way as BindingDB data.

\noindent{\bf KIBA}~\cite{tang2014making} dataset includes kinase inhibitor bioactivities measured in three metrics, $K_i$, $K_d$, and IC$_{50}$. KIBA scores were constructed to optimize the consistency between $K_i$, $K_d$, and IC$_{50}$ by using the statistical information embedded in these quantities~\cite{ozturk2018deepdta}.
The final KIBA dataset we used has 118,254 DTA pairs with 2,068 unique drugs and 229 unique targets. We also randomly split 70\%/10\%/20\% as train/valid/test dataset as in DeepPurpose~\cite{huang2020deeppurpose}.\\
\newline
\noindent{\bf{Unlabeled Molecule and Protein Datasets}}
The unlabeled molecule and protein datasets are from PubChem and Pfam respectively. Details are as follows. \\
\newline
\noindent{\bf Pfam Dataset}~\cite{mistry2021pfam} is a database\footnote{\url{http://pfam.xfam.org/}} of protein families that includes their annotations and multiple sequence alignments generated using hidden Markov models. We use the amino acid sequence of protein extracted from the Pfam database as our unlabeled protein data. The training set we randomly sampled consists of $10M$ protein sequences\footnote{We also study other sizes in Appendix.}.

\noindent{\bf PubChem Dataset}~\cite{kim2016pubchem} is the largest collection of freely accessible chemical information. We use Isomeric SMILES of molecules extracted from PubChem database\footnote{\url{https://pubchem.ncbi.nlm.nih.gov/}} as our unlabeled molecule data. The training set we randomly sampled also consists of $10M$ molecules, which is consistent with unlabeled protein sequences.

\subsection{Model Configurations}
We use two Transformer encoders for molecule encoder $\mathcal{M_D}$ and protein encoder $\mathcal{M_T}$, and each follows RoBERTa$_{\texttt{base}}$ architecture that consists of 12 layers.
The embedding/hidden size and the dimension of feed-forward layer are 768 and 3,072 respectively. The max lengths for molecule and protein are 512 and 1,024 respectively. 
The regression prediction head is 2-MLP layers with \texttt{tanh} activation function and the hidden dimension is 768. 
In general, the model has 178M parameters.
The inference speed is 0.0061s/sample.
Our model implementation is based on Fairseq~\cite{ott2019fairseq} toolkit with version 0.10.2\footnote{\url{https://github.com/pytorch/fairseq/tree/v0.10.2}}.

\subsection{Training and Evaluation}
\noindent{\bf Training.} Our model is optimized by Adam~\cite{kingma2014adam} algorithm with learning rate $1e^{-4}$.
The weight decay is 0.01. The dropout and attention dropout of two encoders is 0.1. The learning rate is warmed up in the first 10k update steps and then linearly decayed. The batch size is 32 sentences and we accumulated the gradients 8 times during training. The maximal training step is 200k. We set the optimal coefficients $\alpha$ and $\beta$ to be 2.0. 

\noindent{\bf Evaluation.} We use (i) mean square error(MSE), (ii) root mean square error(RMSE), (iii) pearson correlation coefficient (PC)~\cite{abbasi2020deepcda}, (iv) concordance index (CI)~\cite{gonen2005concordance} to evaluate the performance of our model on DTA regression task. To have a fair comparison with previous works, the results on the BindingDB dataset are evaluated on (ii) and (iii), and the results on the DAVIS and KIBA datasets are evaluated on (i) and (iv).

\subsection{Compared Baselines}
We compare our SMT-DTA with the following baselines. We focus on state-of-the-art deep learning models as they have demonstrated superior performance over other methods.
\begin{itemize}
    \item \textbf{SAGDTA} uses self-attention mechanisms on the molecular graph to get drug representations. Global and hierarchical pooling are evaluated on DAVIS and KIBA datasets.
    \item \textbf{MGraphDTA} uses a multiscale graph neural network inspired by chemical intuition for DTA prediction, incorporating a dense connection with 27 graph convolutional layers. MGraphDTA was evaluated on DAVIS and KIBA datasets.
    \item \label{DeepDTA_intro}\textbf{DeepDTA}~\cite{ozturk2018deepdta} uses CNN~\cite{albawi2017understanding} on both SMILES and protein sequence to extract their features.
    Originally DeepDTA was evaluated on DAVIS and KIBA datasets, and MONN~\cite{li2020monn} evaluated DeepDTA on the BindingDB dataset.
    \item \textbf{DeepAffinity}~\cite{karimi2019deepaffinity} uses RNN~\cite{medsker2001recurrent} on both SMILES and protein sequence for unsupervised pre-training to learn their representations. After that, CNN layers are appended after RNN for both molecules and proteins to make predictions. DeepAffinity was evaluated on the BindingDB dataset only.  
    \item \textbf{MONN}~\cite{li2020monn} uses a GCN module and a CNN module to extract the features for molecule and protein, respectively. Then, a pairwise interaction module is introduced to link the molecule and protein. MONN was evaluated on the BindingDB dataset. 
    \item \textbf{BACPI}~\cite{li2022bacpi} uses GAT~\cite{velickovic2017graph} to encode molecule graph and CNN to encode protein sequence, respectively. A bi-directional attention is then introduced and the final integrated features are used to make affinity prediction. BACPI was evaluated on the BindingDB dataset.
    \item \textbf{KronRLS}~\cite{pahikkala2015toward} employs the Kronecker Regularized Least Squares (KronRLS) algorithm that utilizes 2D compound similarity-based representation of drugs and Smith-Waterman similarity-based representation of targets~\cite{pahikkala2015toward}. KronRLS was evaluated on DAVIS and KIBA datasets.
    \item \textbf{GraphDTA}~\cite{nguyen2021graphdta} uses GCN~\cite{kipf2016semi}, GAT~\cite{velickovic2017graph}, GIN~\cite{xu2018powerful} and GAT-GCN to encode molecular graph, and CNN to encode protein sequence. Finally, the concatenated features are passed to feed-forward layers for prediction. GraphDTA was evaluated on DAVIS and KIBA datasets.
    \item \textbf{DeepPurpose}~\cite{huang2020deeppurpose} supports the training of customized DTA prediction models by implementing different molecule/protein encoders and various neural architectures. They were evaluated on DAVIS and KIBA datasets.
\end{itemize}

\section{Encoding Process of Transformer Encoder}
We briefly introduce the encoding process of our drug/molecule encoder $\mathcal{M_D}$ and target/protein encoder $\mathcal{M_T}$. 
Take the encoding of protein as an example. As previously defined, the input is the FASTA sequence of the protein, which consists of a $\texttt{[cls]}_T$ token as the representation of the whole protein and amino acid tokens, e.g., $T = \{\texttt{[cls]}_T,\{t_i\}_{i=1}^{\vert T \vert}\}$, $\vert T \vert$ is the protein length and $t_i$ is the amino acid token. For simplicity, we denote $t_0$ as the $\texttt{[cls]}_T$ and $T = \{t_i\}_{i=0}^{\vert T \vert}$.

The first step is to convert each amino acid into its corresponding embedding vector. This is typically done using an embedding matrix.
\begin{align}
\small
    E(T) &= \{E(t_0), E(t_1),...,E(t_T)\}, \nonumber \\
    E(t_i) &= EmbeddingMatrix[i], \nonumber
\end{align}
where $E(t_i)$ is the embedding vector for amino acid $t_i$.
To account for the position of each amino acid in the sequence, we add a positional encoding to each embedding vector. We denote the positional encoding for position $i$ as $PE(i)$.
\begin{equation}
\small
    E'(t_i) = E(t_i) + PE(i), \nonumber
\end{equation}
where $E'(t_i)$ is the embedding vector with positional encoding for amino acid $t_i$.
Then the multi-head attention is used to allow the model to focus on different parts of the sequence simultaneously. Denote that we have $N$ heads, for each head $n$, we compute
\begin{equation}
\small
    AttnHead_n = Attn(Q_n, K_n, V_n), \nonumber
\end{equation}
where $Q_n = W_{Q_n} E'(T), K_n = W_{K_n} E'(T), V_n = W_{V_n} E'(T)$.
$W_{Q_n}$, $W_{K_n}$, and $W_{V_n}$ are learnable weight matrices for queries, keys, and values respectively for the $n$-th head.
The attention scores are computed as the dot product of the query and key, scaled by the square root of the dimension $d_n$, and then passed through a softmax function.
\begin{equation}
\small
    Attn(Q_n, K_n, V_n) = \texttt{softmax}(\frac{Q_n K_n^T}{\sqrt{d_n}}) V_n, \nonumber
\end{equation}
The output from all heads is concatenated and linearly transformed:
\begin{equation}
\small
    MultiHeadOutput = [AttnHead_1, …, AttnHead_N] W_O, \nonumber
\end{equation}
where $W_O$ is the output matrix.
Each position in the multi-head attention output goes through a feed-forward network (the same one for each position):
\begin{equation}
\small
    FFNOutput_i = FFN(MultiHeadOutput_i). \nonumber
\end{equation}
The above multi-head attention and feed-forward network are stacked multiple times (i.e., number of layers). 
The output from the last feed-forward network is the encoded representation of the input sequence: $H_T = \{h_{\texttt{[cls]}_T}, \{h_{t_i}\}_{i=1}^{\vert T \vert}\}$, where $\texttt{[cls]}_T$ is the whole representation of the protein and $h_{t_i}$ represents the encoded hidden state of amino acid $t_i$.

The same encoding process also applies to the input drug SMILES, from which we get $H_D = \{h_{\texttt{[cls]}_D}, \{h_{d_i}\}_{i=1}^{\vert D \vert}\}$.

\end{appendices}

\end{document}